\documentclass[amsmath,amssymb,aps,prx,twocolumn,footinbib,floatfix]{revtex4-1}

\usepackage{subfigure}
\usepackage{enumitem}
\usepackage{graphicx}
\usepackage{dcolumn}
\usepackage{bm}
\usepackage{mathtools}

\usepackage{tabularx}
\usepackage[overload]{empheq}

\renewcommand{\k}{\bm{k}}
\newcommand{\psid}{\psi^{\dagger}}
\renewcommand{\d}{\mathrm{d}}

\begin{document}

\title{Nematic insulator at charge neutrality in twisted bilayer graphene}

\author{Eric Brillaux}
\author{David Carpentier}%
\author{Andrei A. Fedorenko}%
\author{Lucile Savary}%
\affiliation{\mbox{Univ Lyon, ENS de Lyon, Univ Claude Bernard, CNRS, Laboratoire de Physique, F-69342 Lyon, France}}

\date{\today}

\begin{abstract}
We investigate twisted bilayer graphene  near charge neutrality using a generalized Bistritzer-MacDonald continuum model, accounting for corrugation effects. The 
Fermi velocity 
vanishes for particular twist angles properly reproducing the physics of the celebrated magic angles. Using group representation theory, we identify all contact interaction potentials compatible with the symmetries of the model. This enables us to identify two classes of quartic interactions leading to either the opening of a gap or to nematic ordering. We then implement a renormalization group analysis to study the competition between these interactions for a twist angle approaching the first magic value. This combined group theory-renormalization study 
reveals  
that the proximity to the first magic angle favors the occurrence of a layer-polarized, gapped state with a spatial modulation of interlayer correlations, which we call nematic insulator.
\end{abstract}

\maketitle

\section{\label{sec:intro} Introduction}
Within band theory, a reasonable estimate for the relative strength of the quasiparticles kinetic energy is the ratio between the bandwidth of the conducting bands, and some interaction energy. In normal metals, where the density of states at the Fermi energy is nonzero, the excitations of the Fermi sea largely screen  the Coulomb interactions, which renormalizes their strength downwards in a dramatic way. But when the bands near the Fermi
energy disperse very little, even small interactions can lead to significant, qualitative consequences.

The prominent example of such systems is provided by Landau levels and the associated fractional quantum Hall effects~\cite{laughlin_anomalous_1983,nagaosa_anomalous_2010}. More recently, a different class of materials with vanishing bandwidth was uncovered in twisted bilayer graphene (TBG). When two sheets of graphene are rotated with respect to one another by a small angle of about $1.1^\circ$, 
a large moir\'e pattern forms with several thousands of atoms per unit
cell. Remarkably, for some twist angles---the so-called magic angles---the Fermi velocities of the Dirac cones originating from each layer vanish exactly~\cite{bistritzer_moire_2011}. 
Other phenomena accompany this effect such as a minimal bandwidth and a sizeable band gap 
between the conducting and excited bands, under some conditions~\cite{shallcross_electronic_2010, bistritzer_moire_2011, bistritzer_moire_2011-1,  tarnopolsky_origin_2019}. 

TBG gained considerable momentum after the experimental discovery of correlated insulators at various fillings, with neighboring regions of possibly-unconventional superconductivity~\cite{cao_correlated_2018, cao_unconventional_2018, yankowitz_tuning_2018,balents2020superconductivity}. 
In addition, scanning tunneling microscopy and transport data point
toward ``ferromagnetism'' and an ``anomalous Hall effect'' in TBG, but also in thicker van der Waals heterostructures like trilayer graphene~\cite{chen_signatures_2019,chen_tunable_2019,serlin_intrinsic_2019,xie_spectroscopic_2019}.  
The tunability of TBG through a rich phase diagram by electronic gating also sparked numerous works in new directions. While the origin of the superconductivity~remains unclear~\cite{choi_strong_2018,roy_unconventional_2018,wu2018theory,sharma_superconductivity_2019,gu_antiferromagnetism_2019,wu_identification_2019,angeli_valley_2019,cao_nematicity_2020}, evidence is mounting toward the strongest insulator emerging at charge neutrality---where band theory alone would predict a semimetal--- with a charge gap of around $0.86~$meV in the most angle-homogeneous devices~\cite{lu_superconductors_2019}. 
Other scanning measurements find a three-fold rotation symmetry breaking near the first magic angle at charge neutrality~\cite{jiang_charge_2019}. The aim of this article is to 
unveil the nature of 
this rotation symmetry breaking insulator at charge neutrality close to the first magic angle and to provide methodology to analyze its occurrence.  

The study of interacting phases in systems with vanishing bandwidth is notoriously difficult. 
Our strategy to tackle this challenge in TBG is based on the combination of an algebraic identification of interactions preserving the symmetries of the low energy description of TBG and a renormalization group approach to select the generically favored interaction as the twist angle approaches its first magic value. 
In the present case, an additional obstacle lies in the absence of a simple description of the moiré pattern in TBG, thus inhibiting the use of standard field theories.

Our starting point for the non-interacting continuum model of two twisted layers of graphene accounts for several channels of interlayer hoppings. 
These interlayer hoppings renormalize the Fermi velocity, leading to its
vanishing at magic twist angles. 
We develop a diagrammatic technique 
to compute the Green's function and thereby the magic angles to arbitrary order in the interlayer hopping strength $\alpha$ and
non-perturbatively in the imbalance between different hopping channels $\beta$. 
This non-interacting model is then complemented with interactions.  
By formal group theory considerations, we identify all symmetry-allowed contact or short-ranged interactions. 
To determine the most favorable one, we develop
 a renormalization group (RG) technique. The vanishing 
of the bandwidth at the magic angle leads to a singular behavior of the RG: indeed, any interacting potential, while usually treated in perturbations, now corresponds to a dominant energy scale. 
Moreover the first magic angle is not determined by a specific value of a parameter of the free field theory, but 
through a systematic resummation of interlayer hopping terms. 
To overcome these difficulties, we study the scaling behavior of all interacting potentials as the twist angle is varied. When approaching the first magic angle, we monitor the relevance in the RG sense of all potentials, thereby identifying the dominant interacting instability.
We find that as the twist angle approaches the first magic value, a state with both a gap opening and a periodic modulation of interlayer correlations is favored. 
We call this phase a \emph{nematic insulator}.

\begin{figure}[t!]
\includegraphics[scale=1]{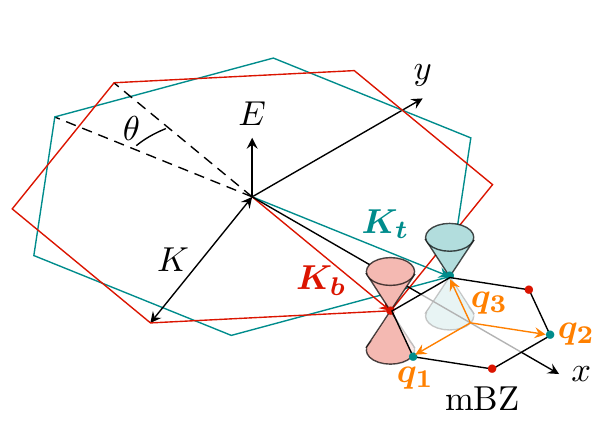}
\caption{%
In a continuum model, the relative twist of the top (green) and bottom (red) layers by an angle $\theta$ leads to one mini-Brillouin zone (mBZ) for each valley of monolayer graphene. The two Dirac cones of the same valley, $\bm{K}_t$ and $\bm{K}_b$, set the sides the mBZ of size $2 K \sin(\theta/2)$, where  $K=|\bm{K}_{t,b}|$ 
is the Dirac momentum of monolayer graphene. 
Electron hoppings between the two layers involve a small momentum transfer
$\bm{q}_j$, $j=1,2,3$ between 
 $\bm{K}_t$ and each of the three nearest $\bm{K}_b$ nodes of the mBZ.
}
\label{fig:BZ_mBZ} 
\end{figure}

\section{\label{sec:model} Free electron Model} 
Following the seminal work of Ref.~\cite{lopesdossantos_graphene_2007}, we treat TBG as a 
periodic moir\'e superlattice characterized by a twist angle~$\theta$. The top and bottom Dirac cones of the 
same valley, denoted $\bm{K}_t$ and $\bm{K}_b$, delineate the mini-Brillouin zone (mBZ) of the superlattice 
(Fig.~\ref{fig:BZ_mBZ}). Focusing on the low energy and long wavelength 
description of TGB, we restrict ourselves to small momentum 
transfers that are diagonal in valley, and thus occur within a single mBZ~\footnote{As standard, we keep 
only the linear part of the  dispersion relation, neglect the $\pm \theta/2$ twists of the wavevectors near the cones and spin-orbit coupling.}. 
The characteristic kinetic energy scale of the model, set by the typical difference of kinetic energy of electrons in 
different layers, is $E_{c} = 2 v_0 K \sin(\theta/2)$, 
where $v_0$ and $K$ are respectively the Fermi velocity and the Dirac momentum of monolayer graphene.
In addition to the kinetic energy in each layer, the single-particle Hamiltonian involves two different interlayer hopping amplitudes.
First, the amplitude $w_1$ of interlayer hopping that
is off-diagonal in graphene sublattice is typically of order 
$w_1 \approx 110$~meV~\cite{lopesdossantos_graphene_2007,kuzmenko_determination_2009}. 
Its strength relative to the kinetic energy is measured by the dimensionless parameter 
$\alpha=w_1/E_{c}$.  
Second, the amplitude $w_2 = \beta w_1$ 
of interlayer hopping that
is diagonal in graphene sublattice is measured by the relative strength $\beta\in[0,1]$ in comparison to off-diagonal hopping.
This relative strength is difficult to determine precisely in experiments, being affected by corrugation 
effects, with typical values evaluated as
$\beta \approx 0.82$~\cite{koshino_maximally_2018, lucignano_crucial_2019}. 
Here we keep $\beta$ as a free parameter. 
Note that our model thus interpolates between the 
Bistritzer-MacDonald continuum (BMC) model~for $\beta=1$~\cite{bistritzer_moire_2011}
and a chirally symmetric continuum (CSC) model~for $\beta=0$~\cite{tarnopolsky_origin_2019}.

Following Ref.~\cite{bistritzer_moire_2011}, we use a rotated basis where the Dirac cones $\bm{K}_{t,b}$ 
of the two layers have the {\em same} $(k_x,k_y)$ coordinates in the mBZ, and measure all energies in units 
of $E_{c}$ (see Appendix~\ref{sec:changes-bases} for details).
The effective Hamiltonian then reads  $H_0' =  H_0+H_\alpha $ with
\begin{equation}\label{eq:single_hamilt} 
H_0= i \left(\bm{\sigma}  \cdot  \bm{\partial} \right) \tau_0,\quad 
H_\alpha=\alpha \sum_{j=1}^3 e^{-i \bm{q_j}\cdot\bm{r}} T_j^+ + \text{h.c.},
\end{equation}
where $\bm{\partial} = (\partial_x,\partial_y)$ and the hopping matrices $T_j^+$
are 
\begin{equation}\label{eq:transfer_matrix} 
T_j^+ = 
\left(\beta \,\sigma_0 + e^{i(j-1)2\pi/3} \sigma_+ +  e^{-i(j-1)2\pi/3} \sigma_-\right) \tau_{+}.
\end{equation}
Here we introduced two sets of Pauli matrices, 
$\sigma$ and $\tau$, which describe respectively the sublattice and layer 
sectors, with $\sigma_z=\pm1={\rm A/B}$ and $\tau_z=\pm1={\rm top/bottom}$.

\begin{figure}[t!]
\centering
\newcommand{\ScaleFree}{0.65}
\subfigure[]{
\includegraphics[scale=\ScaleFree]{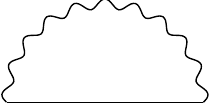}}\hfill
\subfigure[]{
\includegraphics[scale=\ScaleFree]{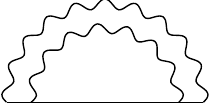}}\hfill
\subfigure[]{
\includegraphics[scale=\ScaleFree]{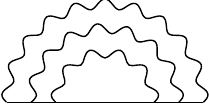}}\hfill
\subfigure[]{
\includegraphics[scale=\ScaleFree]{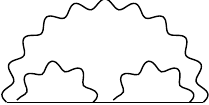}}\hfill
\subfigure[]{
\includegraphics[scale=\ScaleFree]{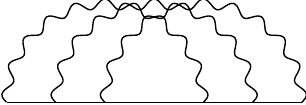}}\hfill
\caption{Diagrammatic expansion of the electron self-energy to order $6$ in the interlayer hopping amplitude
$\alpha=w_1/E_c$ relative to the kinetic energy $E_c$. 
The wavy line represents a pair of opposite hopping processes, summed over all channels with a transfer of momentum $\pm \bm{q}_j$, $j=1,2,3$. 
Diagram~(a) is of order $\alpha^2$, diagram (b) of order $\alpha^4$, and diagrams (c)-(e) are of order $\alpha^6$.
The expansion is non perturbative in the relative strength $\beta$ between hoppings off-diagonal and diagonal in
 sublattices.
}
\label{fig:self_energy}
\end{figure}

\begin{figure}[b!]
    \centering
    \includegraphics[scale=0.54]{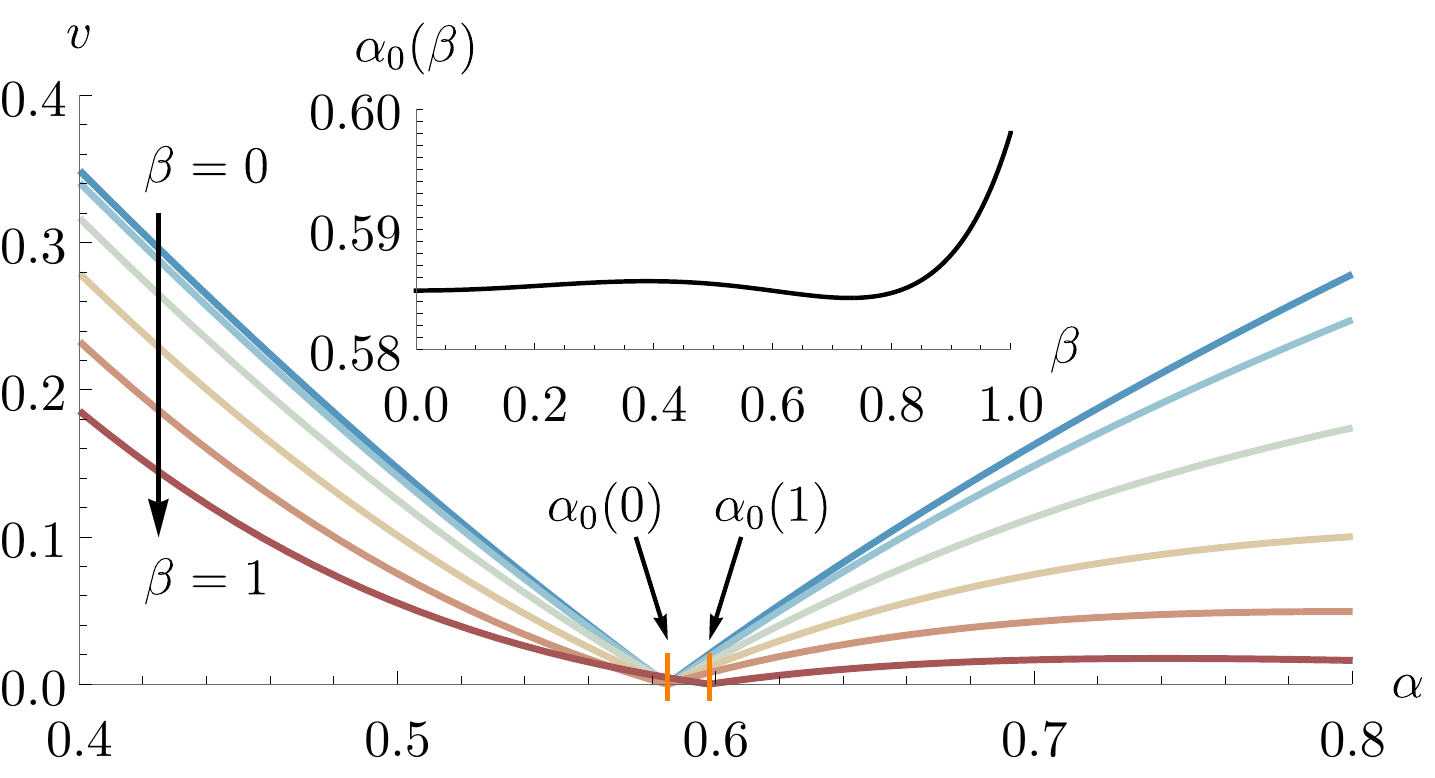}
    \caption{Fermi velocity $v(\alpha,\beta)$ renormalized by interlayer hoppings at order $\alpha^6$, 
    as a function of the relative strength $\alpha$ of the hopping amplitude with respect to the kinetic energy. As the 
    twist angle increases, so does $\alpha$, and the renormalized velocity vanishes at the first magic angle encoded in the first
    magic value $\alpha_0(\beta)$ where $\beta$ sets the asymmetry between diagonal and off-diagonal in sublattice hoppings. 
    Inset: 
    $\alpha_0(\beta)$ depends weakly 
    on corrugation effects, {\it i.e.} on the value of $\beta$. }
    \label{fig:Fermi_velocity}
\end{figure}

The low-energy physics of this model is nontrivial even without interactions. Indeed, the interlayer 
couplings prohibit diagonalizing $H_0'$. This forbids the use of a simple effective 
theory valid for all twisting angles $\theta$ in the vicinity of the magic values. 
As a result, we resort to a free electron model in which interlayer hopping effects are accounted for by a self-energy
which is calculated in a perturbative expansion in $\alpha$. 
Denoting $G_0'$ and $\Sigma$ the translationally-invariant components of the propagator corrected by 
interlayer hoppings and the self-energy respectively, we have
$(G_0')^{-1} = H_0 -  
\partial_\tau - \Sigma 
\approx N_\psi [v(\alpha,\beta) i (\bm{\sigma} \cdot \bm{\partial}) \tau_0 - 
\partial_\tau]$,
where $\partial_\tau$ represents the partial derivative with respect to imaginary time, 
$N_{\psi}$ is a wavefunction normalization 
and $v(\alpha,\beta)$ the Fermi velocity renormalized by the 
hopping processes (see Appendix~\ref{sec:diagrams_nonint}). 
An expansion to order $6$ in $\alpha$ but exact in $\beta$, 
diagrammatically represented in Fig.~\ref{fig:self_energy}, leads~to 
\begin{multline}\label{eq:fermi_velocity} 
N_{\psi}v(\alpha,\beta)=1-3\alpha^2+\alpha^4\left(1-\beta^2\right)^2 
    \\ 
    - \frac{3}{49}\alpha^6\left(37-112 \beta^2 +119 \beta^4 -70 \beta^6\right).
\end{multline}
We call $\alpha_0(\beta)$ the lowest value of $\alpha$ for which this Fermi velocity 
vanishes, which sets the first magic angle value to be approximately $1.1^\circ$.
As shown in Fig.~\ref{fig:Fermi_velocity}, 
this first magic value depends weakly on the parameter $\beta$, and thus on corrugation, and ranges from 
$\alpha_0(1) = 0.598$ for the BMC model to 
$\alpha_0(0)=0.585$ for the CSC model. 
These constitute our first results.

\section{\label{sec:interactions} Symmetry-allowed interactions}
We now identify all short-ranged interaction potentials allowed by the symmetries of the model. 
In order to do so we turn to a field theoretic formalism and consider the Euclidean action,  
$S= S'_0+S_{\rm int}$, 
written as a sum of the free electron term  
\begin{equation}\label{eq:S0prime}
    S'_0 = \int \d^2r \, \d \tau \, \psid(H_0'-\partial_\tau)\psi,
\end{equation}
and an interaction term $S_{\rm int}$ which includes 
generic local quartic couplings between the fermionic fields $\psid$ and $\psi$. 
Using group theoretic methods, detailed in Appendix~\ref{app:symmetries}, 
we identify all 
couplings allowed by the symmetries of the low energy model~\eqref{eq:single_hamilt}~\footnote{%
	Note that in (\ref{eq:inter_action}) the interactions off 
	diagonal in layer acquire an $\mathbf{r}$ dependence as a consequence of our choice of coordinates for 
	the fields in Eq.~\eqref{eq:single_hamilt}
	}. 
This amounts to identiyfing scalar invariants built as 
direct products of irreducible representations of the corresponding symmetry group. 
We find that the allowed couplings are 
(i)  $8$ channels originating from one-dimensional ($1$d)  corepresentations; 
(ii) $4$ channels originating from $2$d corepresentations: 
\begin{multline}
\label{eq:inter_action}
S_{\text{int}} = 
    - \sum_{i=1}^{8} g_i  \int \d^2r \, \d \tau \, 
     \rho^{(i)} (\bm{r}) \rho^{(i)} (\bm{r})
\\
    - \sum_{j=1}^{4} \lambda_j  \int \d^2r \, \d \tau \, 
     \bm{J}^{(j)} (\bm{r}) \cdot \bm{J}^{(j)} (\bm{r}),
\end{multline}
where the densities
$\rho^{(i)} (\bm{r})= \psid R^{(i)}(\bm{r}) \psi$ 
and currents
$\bm{J}^{(j)} (\bm{r})= \psid \bm{M}^{(j)}(\bm{r}) \psi$
involve
coupling matrices $R^{(i)}(\bm{r})$ and $\bm{M}^{(j)}(\bm{r})$.
Following our choice of coordinates for the fields in Eq.~\eqref{eq:single_hamilt}, 
 the coupling matrices in the rotated basis, which enter Eq.~\eqref{eq:inter_action}, are obtained through
$R^{(i)}=A(\bm{r})\hat{R}^{(i)} A^\dagger(\bm{r})$ and
$\bm{M}^{(i)}=A(\bm{r})\hat{\bm{M}}^{(i)} A^\dagger(\bm{r})$, 
 where $A(\mathbf{r})$ is the transformation matrix of the field (see Appendix~\ref{sec:changes-bases}). 
 The coupling matrices $\hat{R}^{(i)} $ and 
 $\hat{\bm{M}}^{(i)}$ are provided in 
Tab.~\ref{tab:matrix_rep}. 
The couplings $g_i$ and $\lambda_j$ are the amplitudes associated with the corresponding coupling potentials. 

\begin{table}[t!]
\renewcommand{\arraystretch}{1.5}
\begin{tabularx}{0.5\textwidth-0.5\columnsep}{Xcccccccc}
\hline
\hline
Corep. &  $~\, A_1^+ ~\,$ &  $~\, a_1^+ ~\,$ & $~\, A_2^+ ~\,$ &  $~\, a_2^+ ~\,$  &  $~\, ~A_1^-~\,$ &  $~\, a_1^-~\,$ & $~\, ~A_2^-\, ~$ &  $\, ~a_2^-~\,$\\
\hline
$ \hat{R}^{(i)} $ & $\sigma_0 \tau_0$ & $\sigma_0 \tau_x$  & $\sigma_0 \tau_z$ & $\sigma_0 \tau_y$ &  $\sigma_z \tau_y$ & $\sigma_z \tau_z$  & $\sigma_z \tau_x$ & $\sigma_z \tau_0$  \\
\hline
$IT$ & \checkmark  & \checkmark & \checkmark & \checkmark &  &  &  & \\
\hline
$C_2$ &  \checkmark & \checkmark &  &  &  \checkmark & \checkmark & &  \\
\hline
$P$ &  & \checkmark &  & \checkmark & & \checkmark & & \checkmark \\
\hline
\hline
\end{tabularx}
\offinterlineskip
\begin{tabularx}{0.5\textwidth-0.5\columnsep}{XXXXc}
Corep. & $E_2^+$ &  $E_4^+$ & $E_2^-$ &  $E_4^-$ \\
\hline
$\sqrt{2}\,\hat{\bm{M}}^{(j)}$ & 
$\bm{\sigma}\tau_0$ & 
$\bm{\sigma}\tau_x$ & 
$\bm{\sigma}\tau_y$ & 
$\bm{\sigma}\tau_z$ \\
\hline
\hline
\end{tabularx}
\caption{
One-dimensional (top) and two-dimensional (bottom) corepresentations (corep.) of the magnetic symmetry group of the continuum model, with their associated coupling matrices $\hat{R}^{(i)}$ and $\hat{\bm{M}}^{(j)}$
expressed in terms of the Pauli matrices in sublattice ($\sigma$) and layer ($\tau$) subspaces. 
 These coupling matrices are normalized such that 
$\text{Tr}[\hat{\bm{M}}^{(j)} \cdot (\hat{\bm{M}}^{(j)})^\dagger]  = 
\text{Tr}[\hat{R}^{(i)} \cdot (\hat{R}^{(i)})^\dagger] = 
4$.
Each one-dimensional corep. can either preserve (\checkmark) or break the combination of inversion and time reversal symmetries $IT$, 
the mirror symmetry $C_2$, and the particle-hole antisymmetry $P$, while preserving the three-fold rotational 
symmetry $C_3$. 
The $\pm$ exponents label the 
eigenvalue of the $IT$ symmetry.
}
\label{tab:matrix_rep} 
\renewcommand{\arraystretch}{1}
\end{table}

The moir\'e pattern is invariant under four discrete ``symmetries'': 
(i) the $2\pi/3$ rotation $C_3 = e^{2i\pi/3 \sigma_z} \tau_0$ around the $z$ axis orthogonal to the bilayer 
    together with 
(ii) the $\pi$ rotation $C_2=\sigma_x \tau_x$ around the $x$ axis of Fig.~\ref{fig:BZ_mBZ} generate the point       group $D_3$, 
(iii) the composition $IT=\sigma_x \tau_0 \mathcal{K}$ of inversion and time reversal 
    is an antiunitary symmetry, where $\mathcal{K}$ denotes complex conjugation, 
and (iv) the unitary particle-hole antisymmetry~\footnote{%
		This antisymmetry is lost when the angular dependence of the kinetic terms 
		$\bm{\sigma}_{\pm\theta/2} \cdot \bm{k}$ is kept, when terms quadratic in momentum are included in the 
		single-particle Hamiltonian, or when intervalley scattering is permitted~\cite{song_all_2018}.
		} 
    $P=\sigma_x \tau_z$, which satisfies $\{P,H_0'\}=0$~\cite{hejazi_multiple_2019}.
The group generated by $D_3$ and $P$ comprises all unitary operations that leave the Hamiltonian invariant up to a sign. We refer to this ensemble as the \emph{unitary} group~$\tilde{D}_3$ of the model. 
It can be decomposed into the semi-direct product 
$\tilde{D}_3 = \{e,P,\bar{e},\bar{P}\} \rtimes  D_3$, where $e=\sigma_0 \tau_0$ is the identity operation, 
$\bar{e}=(PC_2)^2=-e$ and $\bar{P}=\bar{e}P$. The dichromatic magnetic group $\mathcal{M}$ generated by $\tilde{D}_3$ 
and $IT$ can be written as the direct product $\mathcal{M} = \tilde{D}_3 \times \{e,IT\}$. Using the Schur-Frobenius 
criterion~\cite{chen2002group,Ma:2007,Zhong:2004,Atkins:1970,Dresselhaus:2007,Woit:2017}, we 
determine the corepresentations (corep.) of $\mathcal{M}$ from the irreducible representations 
of $\tilde{D}_3$, which can be constructed from 
that of $D_3$ by induction and basic properties of linear representation theory (see Appendix~\ref{app:symmetries}). 

We combine the resulting coupling matrices of Tab.~\ref{tab:matrix_rep} into three sets. 
The eight interactions originating from $1$d coreps.~correspond to 
the density-density couplings diagonal in sublattice while preserving 
$C_3$.
Out of these eight couplings,  
(i) the four interactions associated with $1$d coreps.~which preserves $IT$ are those symmetric on the 
    A/B sublattices, of the form $(\psid \sigma_0 \tau_\mu \psi) 
 	(\psid \sigma_0 \tau_\mu \psi)$ for $\mu =0,x,y,z$~\footnote{Note that we neglected the $\mathbf{r}$ dependence in these expressions for the 
 	sake of clarity.}. They are distinguished by 
 	their breaking of $C_2$ or $P$ symmetries. 
(ii) The four interactions associated with $1$d coreps.~which break $IT$ are those which are antisymmetric in the 
    A/B sublattices, with couplings of the form 
 	$(\psid \sigma_z \tau_\mu \psi) 
 	(\psid \sigma_z \tau_\mu \psi)$. Similarly to set (i), they can break 
 	 $C_2$ or $P$. 
Finally, the (iii) four interactions originating from $2$d coreps.~are current-current couplings between the 
layers, off-diagonal in sublattices, of the form 
$(\psid \bm{\sigma} \tau_\mu \psi) \cdot (\psid \bm{\sigma} \tau_\mu \psi)$.
They break all symmetries of the free model, and in particular the three-fold rotational symmetry $C_3$.

\begin{figure}[t!]
\centering
\subfigure{\label{fig:Relation_Dispersion_gap}}
\subfigure{\label{fig:Relation_Dispersion_G}}
\subfigure{\label{fig:Density_0}}
\subfigure{\label{fig:Density_G}}
\includegraphics[scale=1]{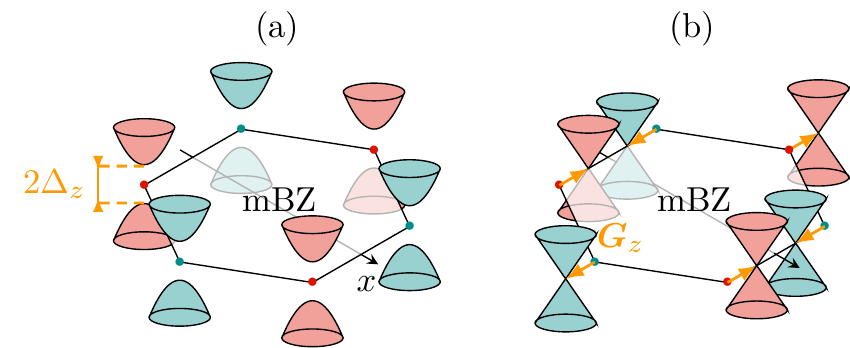}\\[0.3cm]
\includegraphics[scale=0.067]{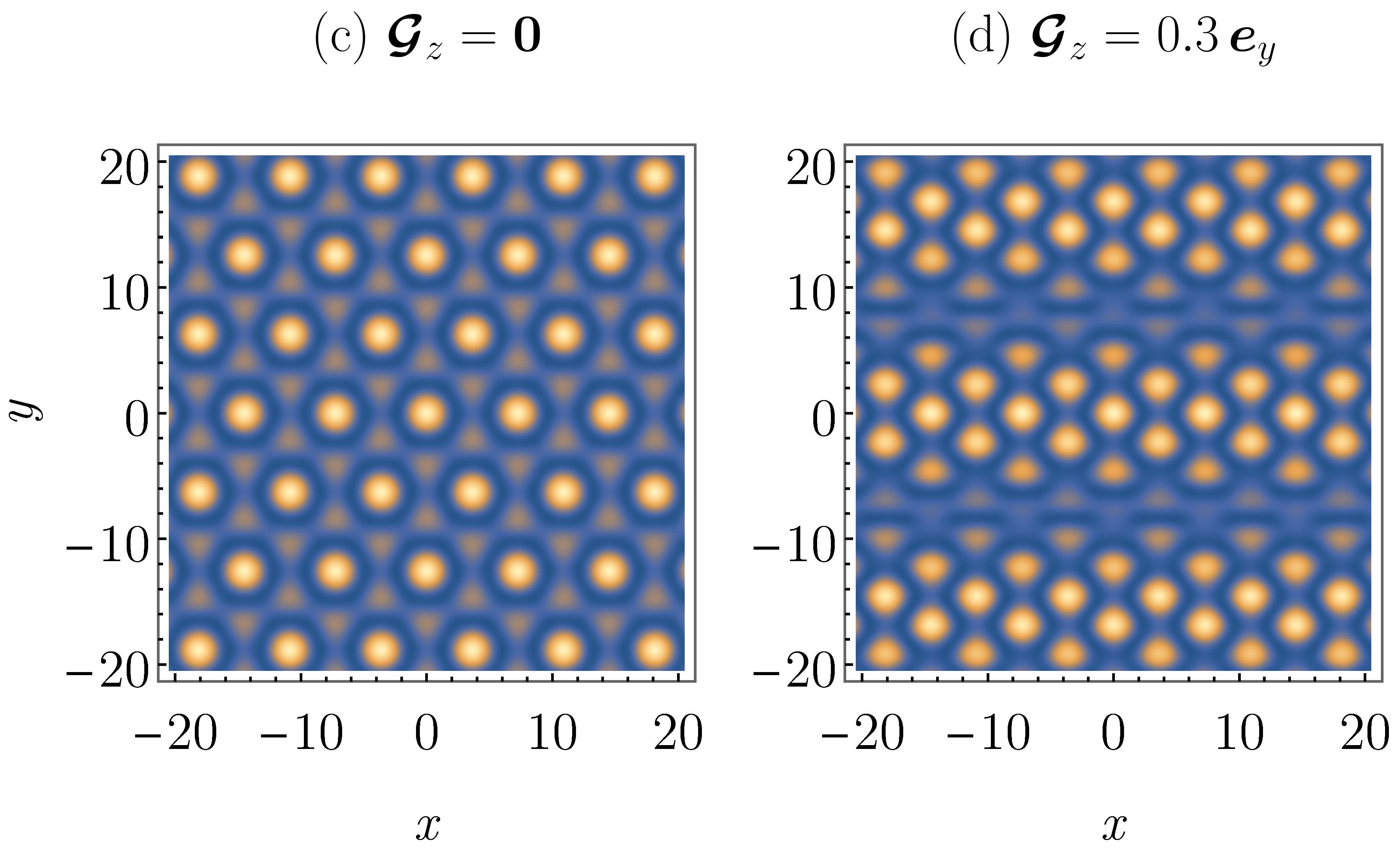}
\caption{
Schematic dispersion relation of (a) a gapped layer-polarized correlated phase 
and (b) a density modulated phase.
While the gapped phase is characterized by the amplitude of the gap $\Delta_z$ opening at the 
$\bm{K}_t$ and $\bm{K}_b$ Dirac points, the second phase is characterized 
by 
a shift of these Dirac points of amplitude $\bm{\mathcal{G}}_z$. This shift leads to a modulation of the relative amplitude of wavefunctions between the two layers, revealed in the density $|\psi_t + \psi_b /2 |^2$ probed {\it e.g.} by a STM tip located on the top layer. The behavior of this density 
is shown (c) without any shift and (d) for a shift $\bm{\mathcal{G}}_z = 0.3 \bm{e}_y$. 
The spatial $C_3$-breaking of this phase is clearly manifested by the appearance of stripes for this density, perpendicular to~$\bm{\mathcal{G}}_z$.
}  
\label{fig:schema_dispersions} 
\end{figure}

\section{\label{sec:MeanField}Nature of the correlated phases}
%
Let us first discuss the nature of the phases induced by these couplings. 
The interactions of type (i), symmetric in sublattices, 
neither open a gap at the Dirac point nor induce a density modulation. 
On the other hand, interactions of type (ii) generate phases with a gap 
$\Delta_\mu \propto 
g_\mu  \langle \psid \sigma_z \tau_\mu \psi \rangle$, 
reminiscent of the gap opening in Boron Nitride, see Fig.~\ref{fig:Relation_Dispersion_gap}.
These various gapped phases are distinguished by their layer correlations. 
Current-current interactions of type (iii) lead to  
radically different phases, in which the Dirac cones of the two layers are shifted with respect to each other by a momentum 
$2\bm{\mathcal{G}}_\mu \propto \lambda_\mu \langle \psid \bm{\sigma} \tau_\mu \psi \rangle$, as shown in Fig.~\ref{fig:Relation_Dispersion_G}. 
They generate gapless phases with $C_3$-breaking 
density modulations.
These spatial modulations are detected when probing the electronic density from one side of the bilayer,  which amounts to coupling 
the local probe asymmetrically to the top and bottom wavefunctions, thus scanning some interlayer density of the form
$|\psi_t + r \psi_b  |^2$, where $0<r<1$ is the asymmetry parameter. 
In Fig.~\ref{fig:schema_dispersions} we 
compare the corresponding density for an asymmetry  $r=1/2$ in the absence of any instability in Fig.~\ref{fig:Density_0} with that in the presence of a 
$C_3$-breaking instability in Fig.~\ref{fig:Density_G}. Stripe-like modulations of 
the interlayer-correlated density are readily observed in this last case.

To gain further insight into the behavior of these phases close to the first magic  angle, we now study their mean-field behavior. 
As we will show later using a renormalization group analysis, only four out of the twelve couplings are sensitive to the proximity of the magic angle. 
They correspond to the interaction potentials diagonal in layers, and originate from 
the 
$a_1^-,a_2^-$ coreps.\ of set
(ii) with respective amplitude $g_z$, $g_0$, 
and the 
$E_4^-,E_2^+$ coreps.\ of set (iii) with amplitudes $\lambda_0$, $\lambda_z$. 
The corresponding order parameters  satisfy the self-consistency
equations \footnote{the momentum integrals run over a finite square of
  the size the ultraviolet cut-off $\Lambda$.} 
\begin{subequations}
\begin{align}
\label{eq:gap_self} 
\Delta_{0/z} & = -2 g_{0/z}  
\displaystyle\int
\mathrm{d} \omega \int_\Lambda \dfrac{
\mathrm{d}^{2} q}{(2\pi)^3} \langle \psid_{q,\omega} \sigma_z \tau_{0/z}
\psi_{q,\omega}\rangle,
\\
\bm{\mathcal{G}}_{0/z} & = -2 \lambda_{0/z} 
\displaystyle\int
\mathrm{d} \omega \int_\Lambda \dfrac{
\mathrm{d}^{2} q}{(2\pi)^3} \langle \psid_{q,\omega} \bm{\sigma} \tau_{0/z}
\psi_{q,\omega}\rangle .
\label{eq:momentum_self}
\end{align}
\label{eq:MFselfconsistent}
\end{subequations}
The correlators in Eq.~\eqref{eq:MFselfconsistent}
 are the translationally-invariant parts of statistical averages computed over the Bloch Hamiltonian density 
$H_{\rm MF}' = H_0' 
+ \bm{\sigma} \cdot (\bm{\mathcal{G}}_0 \tau_0 + \bm{\mathcal{G}}_z \tau_z) +  \sigma_z (\Delta_0 \tau_0 +\Delta_z \tau_z )$. 
The corrections by interlayer hoppings of the correlators in Eq.~\eqref{eq:MFselfconsistent} are obtained within a perturbation expansion in $\alpha$. 
Incorporating the hoppings $H_\alpha$ leads to an enhancement of the order parameters 
by factors $N^{(\mathcal{G}/\Delta)}_{0/z}(\alpha,\beta)$, as is the
case for the renormalization of the Fermi velocity; 
 they are calculated diagrammatically to sixth order in $\alpha$ in Appendix~\ref{appendix-mean-field}. 

\begin{figure*}[ht!]
\subfigure[]{
\includegraphics[scale=0.38]{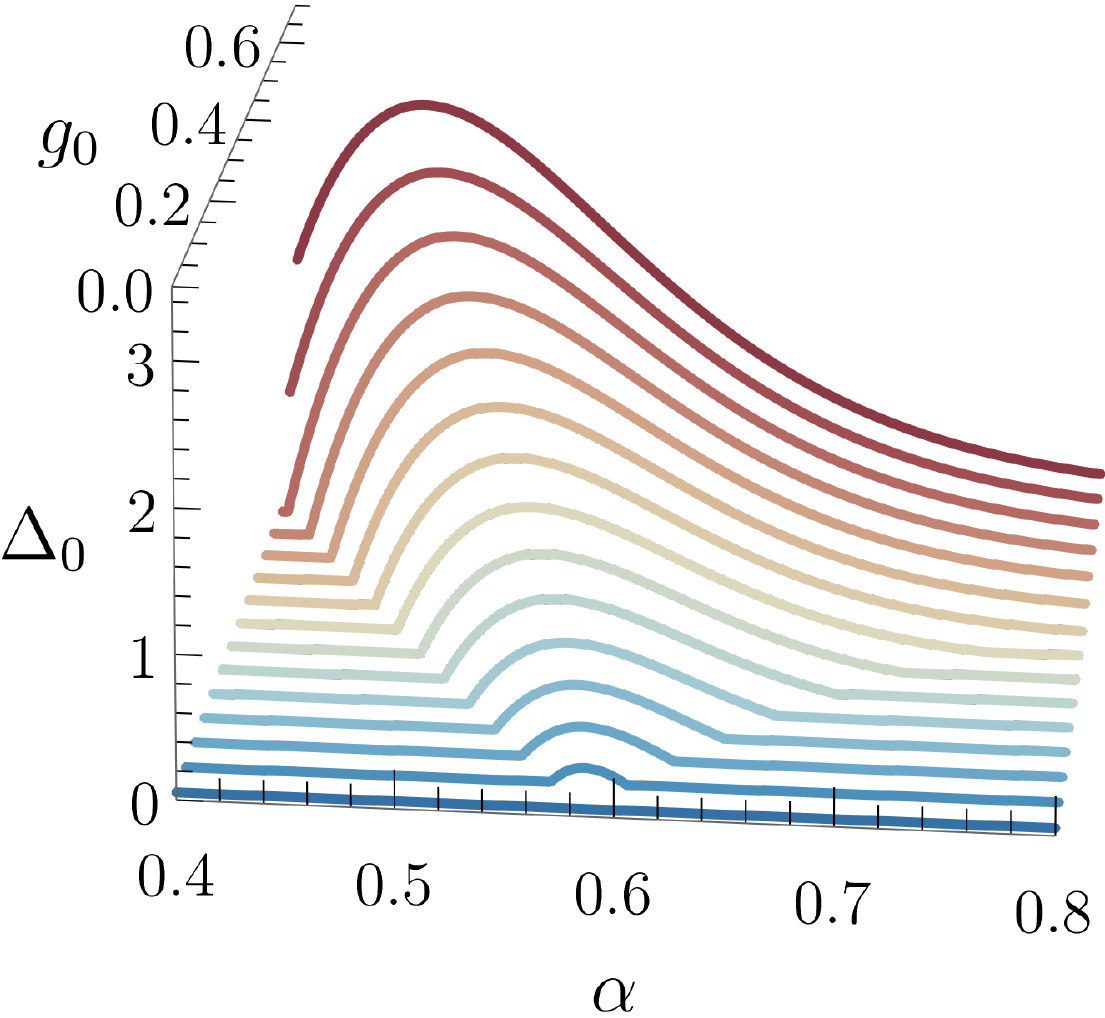}}\hspace{-0.15cm}
\subfigure[]{
\includegraphics[scale=0.38]{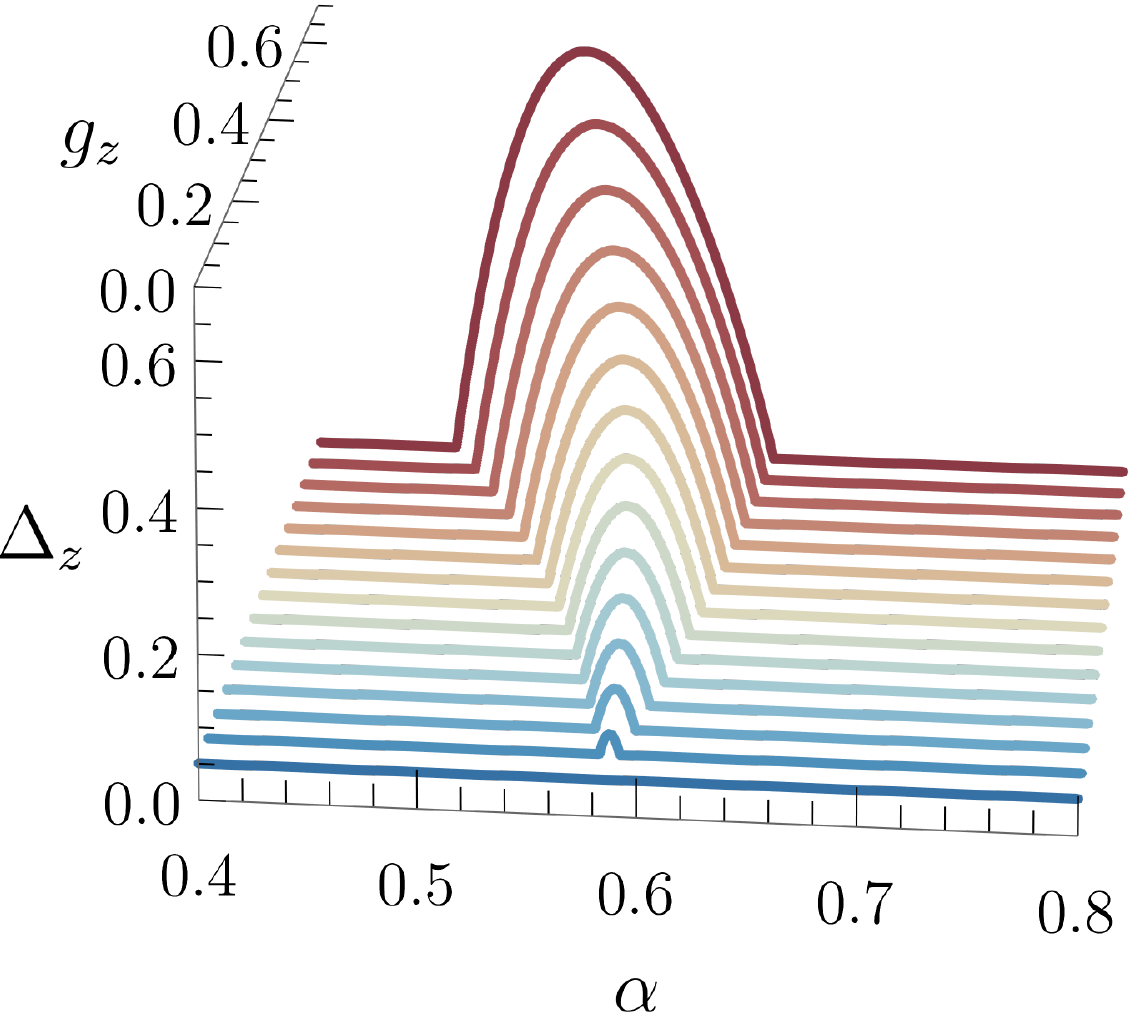}}\hspace{-0.15cm}
\subfigure[]{
\includegraphics[scale=0.38]{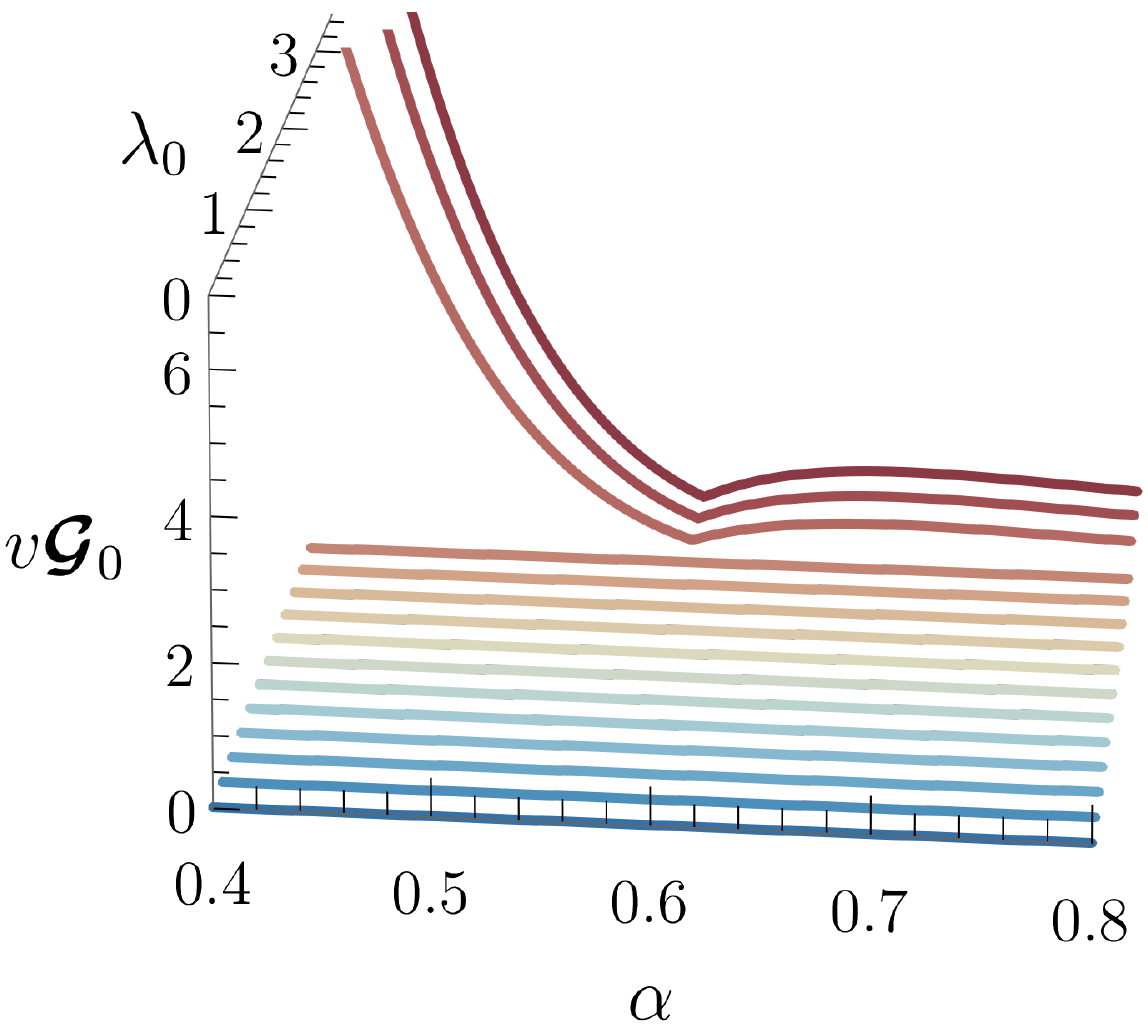}}\hspace{-0.15cm}
\subfigure[]{
\includegraphics[scale=0.38]{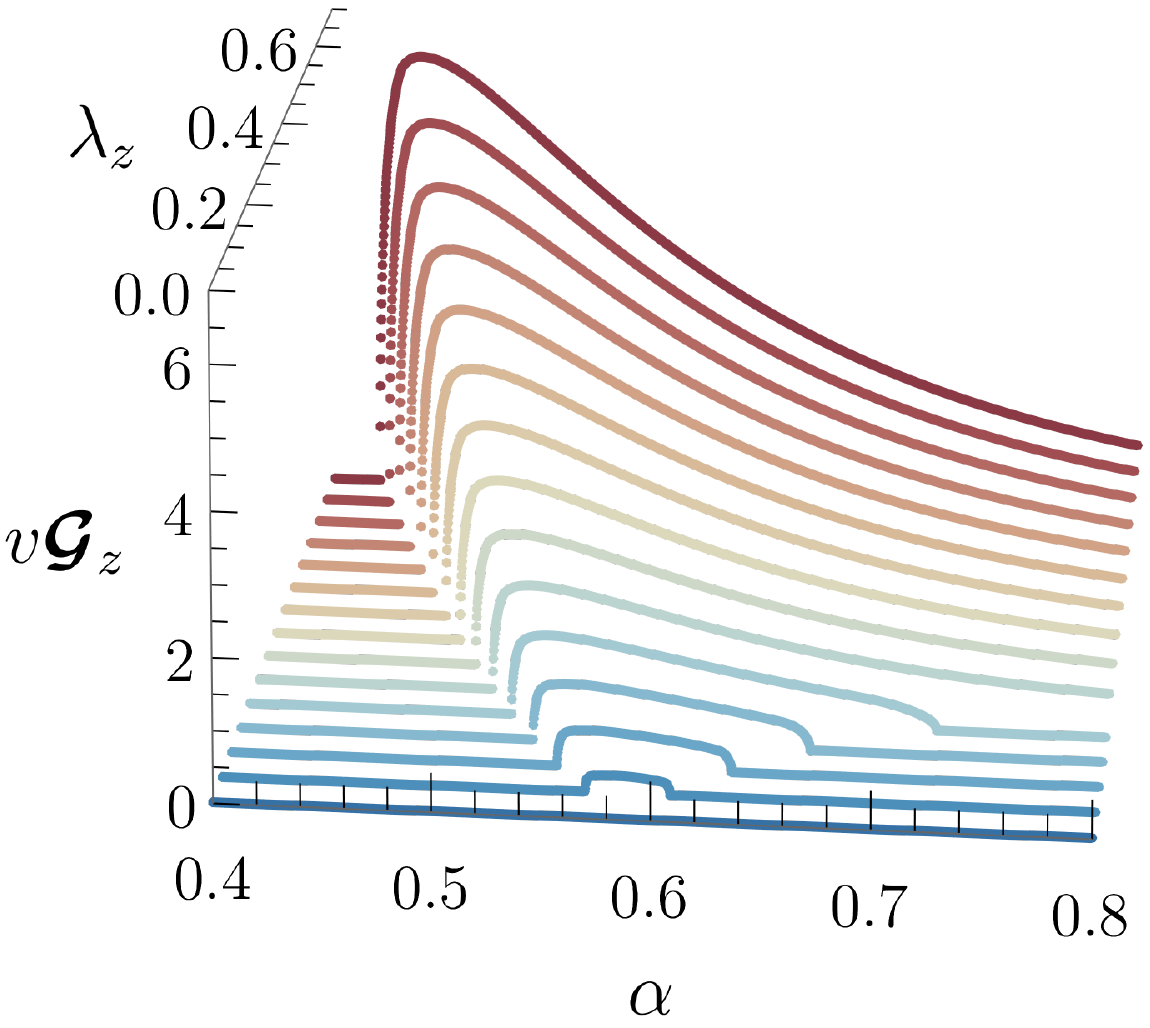}}
\caption{%
Mean-field order parameters of the leading instabilities as a function of the twisting angle encoded in $\alpha$ and for various coupling strengths.  
Layer-polarized gap $\Delta_z$ (a) and $\Delta_0$ (b). (c) and (d) Energy $v\bm{\mathcal{G}}_z$ and $v\bm{\mathcal{G}}_0$ of the 
layer-polarized density modulated phase
associated to shift $\bm{\mathcal{G}}_z$ and $\bm{\mathcal{G}}_0$ of the Dirac
cones (here the shifts are oriented along the $y$ axis). 
The amplitudes of the order parameters are provided for $\beta=0.82$ and 
in arbitrary  units of $10^{-2}\Lambda$ where $\Lambda$ is an energy cut-off.}
\label{fig:order_param_alpha} 
\end{figure*}
The resulting dependence of each separate order parameter on 
the proximity to the magic angle and for various strengths of the couplings is depicted in Fig.~\ref{fig:order_param_alpha}.
The insulating phases, characterized by a gap $\Delta_0$ or $\Delta_z$, develop at a critical coupling which decreases as the parameter $\alpha$ approaches its magic value~$\alpha_0$.
Such gapped phases generically occur 
in some range of twist angles around the magic value, in agreement with the experimental findings of Ref.~\cite{codecido2019correlated}. 
At the mean-field level and for a fixed $\alpha$, we find that a $\Delta_0$ insulator occurs for 
weaker couplings $g_0$ 
than the couplings $g_z$ required for the appearance of the $\Delta_z$ 
insulator. 
We will see that this hierarchy is modified when fluctuations are accounted for, demonstrating the necessity to develop the RG approach. 
The situation for the $C_3$ symmetry breaking phases with periodic modulation is different: while the phase which is antisymmetric in layers, characterized by a $\bm{\mathcal{G}}_{z}$ momentum, is also favored by the vanishing bandwidth close to $\alpha_0$, the finite critical strength for 
the analogous phase symmetric in layer, associated with  $\bm{\mathcal{G}}_{0}$, does not depend on~$\alpha$. 

Having identified all interacting instabilities of TBG and established their strong enhancement close to the magic angle, 
we now study the competition between them by resorting to a renormalization group technique.

\section{\label{sec:interacting_RG} Renormalization group picture}
As we have seen, the vanishing of the kinetic energy scale 
set by the renormalized velocity $v$  
entails that the four non-trivial 
interactions are relevant close to the first magic angle. 
In order to identify the leading instability, we study {\it via} a RG approach 
the competition between them as $\alpha$ approaches the 
first magic value. Our starting point is the field theory described by the action $S= S'_0+S_{\rm int}$ 
where the quadratic action $S'_0$ given in Eq.~\eqref{eq:S0prime}
involves both the kinetic energy and the interlayer hoppings. 
We sum over the latter to capture the vanishing energy scale.
Hence, we expand the correlation functions of the interacting theory not only in the coupling constants but also in the amplitude of interlayer hoppings.

\begin{figure}[b!]
\newcommand{\ScaleInt}{0.95}
\subfigure[
]{\includegraphics[scale=\ScaleInt]{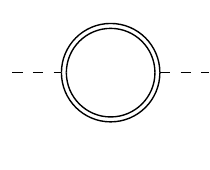}}
\subfigure[
]{\includegraphics[scale=\ScaleInt]{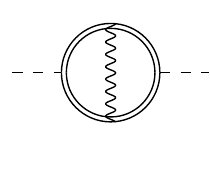}}
\subfigure[
]{\includegraphics[scale=\ScaleInt]{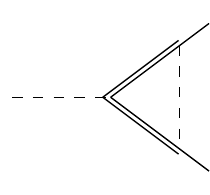}}
\subfigure[
]{\includegraphics[scale=\ScaleInt]{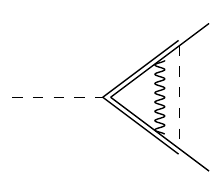}}
\subfigure[
]{\includegraphics[scale=\ScaleInt]{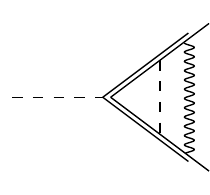}}
\subfigure[
]{\includegraphics[scale=\ScaleInt]{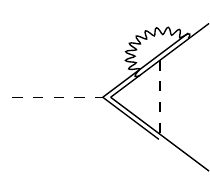}}
\subfigure[
]{\includegraphics[scale=\ScaleInt]{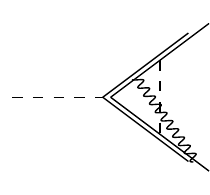}}
\caption{%
(a) - (b) Polarization (bosonic self-energy) to first order in the couplings, (a) at order $\alpha^0$ and (b) at order $\alpha^2$. 
(c) - (g) Three-point vertex to first order in the couplings, (c) at order $\alpha^0$; (d) - (e) at order $\alpha^2$ with multiplicity one; (f) - (g) at order $\alpha^2$ with multiplicity two. The double line is the fermionic propagator corrected by interlayer hoppings, while the dashed line is the bosonic propagator. 
The wavy line represents a pair of opposite hopping processes, summed over all channels with a transfer of momentum $\pm \bm{q}_j$ for $j=1,2,3$.
}
\label{fig:vertex} 
\end{figure}

Motivated by the experimental observations of an insulating behavior at charge neutrality, 
 we carry out a ``particle-hole'' Hubbard-Stratonovich transformation suitable to describe gapped phases (as opposed to superconductors). We introduce
the scalar bosonic fields 
$\phi_i$, $i=1,...,8$ for each 1d channel and the 
vector bosonic fields $\bm{\varphi}_j$, $j=1,...,4$ for each 2d channel,
so that the interaction part of the action~\eqref{eq:inter_action} becomes:
\begin{multline}
  \label{eq:1}
  S_{\rm int}\rightarrow \sum_{i=1}^8 \int \mathrm{d}^{d-1}r\,\mathrm{d}\tau\left(\phi_i^2
    +2\sqrt{g_i}\,\phi_i \psi^\dagger R^{(i)}(\bm{r})\psi\right) \\
  + \sum_{j=1}^4 \mathrm{d}^{d-1}r\,\mathrm{d}\tau\left(\bm{\varphi}_j^2
    +2\sqrt{\lambda_j} \, \bm{\varphi}_j \cdot\psi^\dagger\bm{M}^{(j)}(\bm{r})\psi\right).
\end{multline}
We expand around the lower critical dimension, setting the space-time dimension to $d=2+\epsilon$, 
and renormalize the theory using
the minimal subtraction scheme. 
We introduce the renormalized couplings $g\in\{g_i,\lambda_j\}$ related to the bare ones by $\mathring{g} = \mu^{-\epsilon}N_\psi^2 Z_g^2 Z_{\phi}^{-1}g$. 
Here~$\mu$ is the momentum scale at which we renormalize the theory
and 
$N_\psi$ is the wavefunction normalization factor generated by interlayer hoppings which was introduced previously. 
The renormalization constants $Z_g$ and $Z_\phi$ for $\phi\in\{\phi_i,\bm{\varphi}_j\}$ absorb the poles of the three-point vertex and the bosonic self-energy, respectively.
We expand these functions to first order in the interaction couplings
using the Green's function 
corrected by interlayer hoppings $G_0'$ as the fermionic propagator. This expansion is represented by the diagrams shown in Fig.~\ref{fig:vertex},
where interlayer hoppings are treated perturbatively to second order in~$\alpha$.
We obtain the RG flow equations for the coupling constants  
as a function of the parameters $\alpha$ and $\beta$.
%
Crucially, we find that only four
interactions out of twelve have non-zero divergent corrections. The eight other couplings have a trivial
flow, either because the correction has no pole in $\epsilon$---these correspond to the four
channels with $\sigma_0$ sublattice structure; or because the correction vanishes at low energy 
---these correspond to the four channels that
are off-diagonal in layer space.
We thus restrict our study to the 
four-dimensional subspace corresponding to the instabilities of Tab.~\ref{tab:twoFP_isolated}, which are all
associated with a phase transition toward a correlated phase. These relevant couplings are all diagonal in layer.

\begin{table}[t!]
\begin{center}
\renewcommand{\arraystretch}{2}
\begin{tabularx}{0.5\textwidth-0.5\columnsep}{cccXl}
\hline
\hline
Channel$\quad$ & Coupling~ & $\hat{\bm{M}}_i$  & & FP $g_i^*(\alpha,\beta)$\\
\hline
   $a_2^-$ $\quad$ & $g_0$ & $\sigma_z \tau_0$ & & $\pi v \epsilon/4\left[1 - 12\alpha^2(1-\beta^2)\right]$ \\\hline
   $a_1^-$ $\quad$ & $g_z$ & $\sigma_z \tau_z$ & & $\pi v \epsilon/4$ \\\hline
    $E_2^+$ $\quad$ & $\lambda_0$ & $\bm{\sigma}\tau_0/\sqrt{2}$ & & $\pi v \epsilon/4\left[1-3 \alpha^2(1-\beta^2)\right]$ \\\hline
 $E_4^-$ $\quad$ & $\lambda_z$ & $\bm{\sigma}\tau_z/\sqrt{2}$ & & $\pi v \epsilon/4\left[1 + 3 \alpha ^2(1+\beta^2)\right]$ \\
\hline
\hline
\end{tabularx}
\end{center}
\caption{Isolated, non-gaussian critical fixed points (FPs) for the four non-trivial instabilities.
}
\label{tab:twoFP_isolated}
\end{table}
We now briefly discuss the essential features of this four-dimensional flow.
The gaussian fixed point (FP) at the origin is always stable in $d=3$. 
Besides, we identify four critical points, one for each non-trivial coupling, 
listed in Tab.~\ref{tab:twoFP_isolated}. 
They control phase transitions toward the four correlated phases discussed in the mean-field analysis.  
As $\alpha$ approaches the magic value $\alpha_0$, all four critical FPs collapse towards the gaussian FP. 
Meanwhile, the (Dirac) semimetallic region, which corresponds to the basin of attraction of the gaussian FP, shrinks and disappears completely. As a result,  
these four couplings  
are always relevant close enough to the magic angle, regardless of the value of the bare 
interaction strength.
This scenario provides a natural way of identifying the dominant instabilities near the magic angle: they correspond to the couplings whose critical FPs collapse the fastest towards the origin. 

\begin{figure}[t!]
\centering
\hspace{-0.3cm}
\subfigure{\label{fig:flow_alpha=0} 
\includegraphics[scale=1]{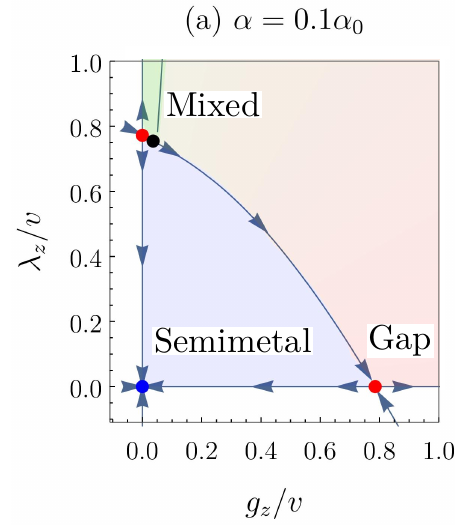}}\hspace{-0.1cm}
\subfigure{\label{fig:flow_alpha=alpha0} 
\includegraphics[scale=1]{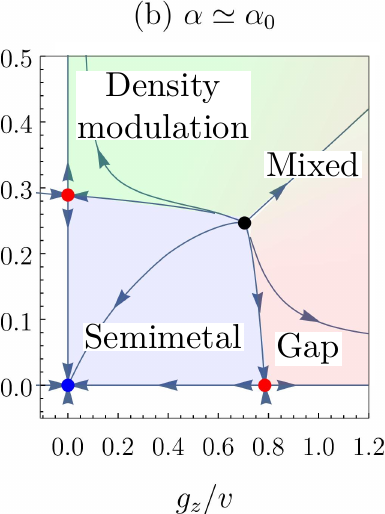}}
\caption{Renormalization flow of the couplings $g_z$ and 
$\lambda_z$ at $\beta = 0.82$ ~\cite{koshino_maximally_2018, lucignano_crucial_2019}, (a) for a weak interlayer hopping amplitude, $\alpha=0.1\alpha_0$; 
and~(b) close to the first magic angle, $\alpha \simeq \alpha_0$. As $\alpha$ approaches $\alpha_0$, the blue semimetal region shrinks to the origin. To study the competition between the couplings we rescale them by the vanishing velocity~$v$. The red critical FPs control the transitions toward the gapped (red region) or density-modulated (green region) phases. 
The black source FP gives rise to a crossover region (mixed state). It migrates away from the vertical axis as we increase $\alpha$, thus expanding the density-modulated region.
}
\label{fig:RGflow} 
\end{figure}

Hence, as follows from Tab.~\ref{tab:twoFP_isolated}, 
we  discard 
the couplings $g_0$ and $\lambda_0$ (the amplitudes of interactions which are  symmetric in layers) and 
focus on the competition between the couplings which are antisymmetric in layers, 
of amplitude $g_z$  
associated with the layer-polarized gapped phase and 
$\lambda_z$ associated with the $C_3$-breaking density-modulated phase. 
We note that, while the gapped phase is reminiscent of the 
dynamical mass generation in the Gross-Neveu model~\cite{classen_mott_2015, rosenstein_dynamical_1991}, the $C_3$-breaking density-modulated 
phase is specific to TBG. 
The competition between the two most relevant instabilities is dictated by the following coupled RG flows (for derivation see Appendix~\ref{sec:diagrams_int})
%
\begin{align}
- \mu \dfrac{\partial g_z}{\partial \mu}  & = - \epsilon g_z +\frac{4
  g_z^2}{\pi  v}  +\frac{4  g_z \lambda_z}{\pi  v} \left[1-6\alpha ^2\left(1-\beta ^2\right)\right], \\
 - \mu \dfrac{\partial \lambda_z}{\partial \mu}  & =   - \epsilon \lambda_z +\frac{4
  \lambda_z^2}{\pi  v} \left[1+3\alpha^2\left(1+\beta^2\right)\right] 
  \nonumber \\ 
  & \quad \quad 
  +\frac{2 \lambda_z g_z}{\pi  v} \left[1-6\alpha ^2\left(1-\beta ^2\right)\right].
 \label{eq:beta_lambda}
\end{align}\label{eq:beta_g}
As mentioned above, all FPs collapse to the origin at the magic angle. Thus to explore the competition between the phases, we plot the renormalization flow 
for the couplings rescaled by the vanishing velocity.
The effect of the proximity to the magic angle on this competition is shown in Fig.~\ref{fig:RGflow}, 
where we compare the flow close to the first magic angle (b) 
with that for the case when interlayer hopping is suppressed (a) \footnote{
As is the case for the magic angle value, 
the renormalization flow weakly depends on corrugation effects, i.e. on $\beta$. The major impact of corrugation is to slow down the shrinking of the semimetallic region by pushing away the crossover point from the origin as $\beta$ decreases from $1$.}.
This comparison shows that the proximity to the magic angle favors the occurrence of density modulations.
The large scale behavior is dominated by the fastest diverging coupling, whether $g_z$ or $\lambda_z$. Within our perturbative RG analysis, a crossover line separates the corresponding regions, whose parametric equation reads $\lambda_z = 
    g_z [1+6\alpha^2(1-\beta^2 )   ]/ [6\alpha^2(3-\beta^2)]
    $.
Around the crossover line, 
both order parameters coexist over a large range of length scales, corresponding to the appearance of a gapped, periodically modulated state, asymmetric in layers and breaking the $C_3$ and $IT$ symmetries. We call it a nematic insulator by analogy with phases discussed in Ref.~\cite{fradkin2010nematic}. 
This nematic insulating behavior is  characterized by a runaway RG flow of both $\lambda_z,g_z$.
It appears for a wide range of coupling parameters, as a consequence of the proximity to the magic angle.

\section{\label{sec:summary} Discussion and outlook}
Applying group theory supplemented by a renormalization group approach,
we found that a gapped nematic state with $C_3$ breaking modulation of density is favored at charge neutrality in TBG when the twist angle approaches its first magic value.  
A gap was observed at charge neutrality in TBG both in scanning tunneling microscopy and spectroscopy studies
\cite{jiang_charge_2019,choi2019electronic,kerelsky2019maximized,xie_spectroscopic_2019} as well as in 
 four-terminal transport measurements \cite{lu_superconductors_2019}. Three-fold symmetry breaking and nematic ordering were
 also reported 
  \cite{jiang_charge_2019,kerelsky2019maximized,cao_nematicity_2020}. 
Both of these experimental observations strongly support the occurence of a nematic insulating state at charge neutrality in TBG, as we obtain within our RG scenario. 
We also find that such a state persists
even when the strength of interactions is weakened by screening as was experimentally observed in Ref.~\cite{stepanov2020untying}. 
Let us  stress that our RG approach identifies a gapped nematic behavior in the perturbative scaling regime, but does not rule out that other types of correlations develop at larger length scales, including those of 
intervalley-coherent and generalized ferromagnetic insulating states recently discussed  in Refs.
\cite{ochi_possible_2018,bultinck2019ground,chichinadze2020orbital,zhang2020correlated}. 
It is worth noting that the energies of these different ground states seem to be very close to each other, suggesting 
a strong sensitivity to experimental conditions: indeed,  h-BN encapsulation, which induces chirality breaking, favors the layer-polarized insulators such as the nematic insulator discussed in this paper as opposed to intervalley-coherent or generalized ferromagnetic insulating states~\cite{kang2019strong,zhang2020correlated}. 
 Finally, let us
 note that the occurence of an analogous nematic insulating state close to quantum spin Hall phase transitions raises the questions of its relation with the topological nature of the underlying semimetal~\cite{hejazi_multiple_2019,song2018all,liu2019nematic}.

\acknowledgments
We would like to thank Leon Balents for valuable discussions.
We acknowledge support from the French Agence Nationale de la
Recherche through Grant No.~ANR-17-CE30-0023 (DIRAC3D), the ToRe
IdexLyon breakthrough program, and funding from the European Research
Council (ERC) under the European Union's Horizon 2020 research and
innovation program (Grant agreement No.~853116, ``TRANSPORT''). 


\appendix

\section{Change of basis}
\label{sec:changes-bases}

The Hamiltonian describing the low-energy physics near the two twisted Dirac cones at $\bm{K}_{t,b}$ originating from a single valley of graphene can be written as~\cite{balents2019general}
\begin{equation}
  \label{eq:supp10}
  \hat{H}=\int \mathrm{d}^2r \, \hat{\psi}^\dagger\begin{pmatrix}
v_0\bm{\sigma}\!\cdot\!\left(i\bm{\partial}+\frac{\bm{q}_1}{2}\right) & \hat{T}^\dagger(\bm{r})\\
\hat{T}(\bm{r}) &
v_0\bm{\sigma}\!\cdot\!\left(i\bm{\partial}-\frac{\bm{q}_1}{2}\right)
\end{pmatrix}
\hat{\psi}.
\end{equation}
The momentum $\bm{q}_1 = \bm{K}_t-\bm{K}_b$ gives 
the relative displacement of the Dirac momentum $\bm{K}$ of each layer due to the twist, while $v_0$ is the Fermi velocity of graphene. Notice that there are three equivalent $\bm{K}$ points in monolayer graphene, each leading to one copy of the Hamiltonian $\hat{H}$ with a relative displacement $\bm{q}_j$, $j=1,2,3$, where the momenta $\bm{q}_2$ and $\bm{q}_3$ are obtained through a rotation of $\bm{q}_1$ by an angle of $2\pi/3$ and $4\pi/3$ respectively. The interlayer hopping matrix $\hat{T}(\bm{r})$ reads
\begin{equation}
\begin{cases}
\hat{T}(\mathbf{r})=\displaystyle\sum_{j=1}^3 e^{-i(\bm{q}_j-\bm{q}_1)\cdot\bm{r}} T_j^+ + \text{h.c.}\\
T_j=\dfrac{t_{\rm AA}}{3}\sigma_0+\dfrac{t_{\rm
  AB}}{3}\left(\sigma_+ e^{-2i(j-1)\pi/3} + {\rm h.c.} \right)
\end{cases},
\end{equation}
where $t_{\rm AA}$ and $t_{\rm AB}$ are the hopping amplitudes in the AA and AB/BA regions, respectively.

Hamiltonian~(\ref{eq:supp10}) is simplified by 
rotating the basis \cite{bistritzer_moire_2011}
\begin{equation}
  \label{eq:supp2}
  \hat{\psi}(\bm{r},\tau) = A_1(\bm{r})\psi (\mathbf{r},\tau), \quad A_j(\bm{r})= e^{-i(\bm{q}_j\cdot\bm{r}/2)\tau_z},
\end{equation}
which brings the Dirac cones to the same momentum:
\begin{equation}
  \label{eq:supp11}
  H=\int \mathrm{d}^2r \,\psi^\dagger \begin{pmatrix}
v_0\bm{\sigma}\cdot i\bm{\partial} & T^\dagger(\bm{r})\\
T(\bm{r}) &
v_0\bm{\sigma}\cdot i\bm{\partial} 
\end{pmatrix}\psi ,
\end{equation}
where $ T(\bm{r})=\sum_{j=1}^3e^{-i\bm{q}_j\cdot\bm{r}} T_j^+$. Applying the same 
change of basis to quartic terms in the action, {\it e.g.}
corresponding to a density-density interaction of the form
\begin{multline}
  \label{eq:supp5}
  S_{\text{int}} =  g \int \mathrm{d}^2 r \, \mathrm{d} \tau \,
                  \hat{\psi}^\dagger(\bm{r},\tau)
                  \hat{R}\hat{\psi}(\bm{r},\tau)
                  \\ \hat{\psi}^\dagger(\bm{r},\tau)
                  \hat{R} \hat{\psi}(\bm{r},\tau),
\end{multline}
we arrive at
\begin{multline}
  \label{eq:supp6}
   S_{\text{int}} =  g \int \mathrm{d}^2 r \, \mathrm{d} \tau \,
                  \psi^\dagger(\bm{r},\tau)
                  R(\bm{r})\psi(\bm{r},\tau)
                  \\ \psi^\dagger(\bm{r},\tau)
                  R(\bm{r}) \psi(\bm{r},\tau)
\end{multline}
with the rotated interaction matrix
\begin{equation}
  \label{eq:supp4}
  R(\bm{r}) = \dfrac{1}{3}\sum_{j=1}^3 A_j^\dagger(\bm{r}) \hat{R} A_j(\bm{r}).
\end{equation}
Though both $\hat{R}$ and $R(\bm{r})$ describe contact interactions,  while $\hat{R}$ is space-independent, $R(\bm{r})$ depends in general on the position as a consequence of Eq.~\eqref{eq:supp4}.

(i) If $\hat{R}$ is diagonal in layer, {\it i.e.} proportionnal to $\tau_{0/z}$, it commutes with $A_j(\bm{r})$ so that
$R(\bm{r})=\hat{R}$.

(ii) If $\hat{R}$ is not diagonal in layer, {\it i.e.} proportionnal to $\tau_{x/y}$, it does not commute with $A_j(\bm{r})$ so that
$R(\bm{r})$ differs from $\hat{R}$. In that case, $R(\bm{r})$ is modulated periodically over a distance of the order of the moire lattice constant. Indeed,  we have
\begin{equation}
\begin{cases}
\frac{1}{3}\displaystyle\sum_{j=1}^3 A_j^\dagger(\bm{r}) \tau_x A_j(\bm{r}) = f_1(\bm{r}) \tau_x + f_2(\bm{r}) \tau_y \\
\frac{1}{3}\displaystyle\sum_{j=1}^3 A_j^\dagger(\bm{r}) \tau_y A_j(\bm{r}) = f_2(\bm{r}) \tau_x + f_1(\bm{r}) \tau_y
\end{cases},
\end{equation}
with $f_1(\bm{r})= \frac{1}{3}\sum_{j=1}^3 \cos(\bm{q}_j\!\cdot\!\bm{r})$, $f_2(\bm{r})= \frac{1}{3}\sum_{j=1}^3 \sin(\bm{q}_j\!\cdot\! \bm{r})$.
These results also apply to current-current quartic interactions, where $\hat{R}$ is replaced by a vector of matrices~$\hat{\bm{M}}$.

\section{\label{sec:diagrams_nonint}Diagrammatic technique for the non-interacting theory}
\setcounter{equation}{0}

\label{sec:non-int_dispersion} 
To expand any observable 
in interlayer hoppings in the absence of interactions, it is not 
 mandatory to resort to a field theoretical approach. We do so
 however, because it is useful for applying 
RG when interactions are included. To that end we need 
to introduce the Feynman rules specific to this unusual field theory. The free fermionic propagator associated to the action of the decoupled bilayer $S_0 = \int \d^2r \, \d \tau \, \psid(H_0-\partial_\tau)\psi$ reads
\begin{equation}\label{eqS:free_propag} 
G_0(\bm{k},\Omega) = (\bm{\sigma} \! \cdot \! \bm{k} -i\Omega)^{-1}
\end{equation}
in Fourier space, where  $\bm{k}$ is the momentum, $\Omega$ the Mastubara frequency, and we omitted the identity matrices $\sigma_0$ and $\tau_0$ for simplicity. When drawing Feynman diagrams, we represent the free propagator~\eqref{eqS:free_propag} with a 
solid line. We reserve 
$\bm{q}$ and $\omega$ for the internal momentum and Matsubara frequency and use $\bm{k}$ and $\Omega$ for external ones. 
Any correlation function can be written as an ensemble average $\langle ... \rangle_0$ over~$S_0$. In particular for the time-ordered two-point function we have
\begin{equation}
\label{eqS:St_expansion} 
\langle \mathcal{T} \psi \psid \rangle_0' = \dfrac{\langle \mathcal{T} \psi \psid \, e^{-S_\alpha} \rangle_0}{\langle e^{-S_\alpha} \rangle_0},
\end{equation}
where $\langle ... \rangle_0'$ denotes the ensemble average over the quadratic action $S_0' = S_0 + S_\alpha$, which includes the hopping action $S_\alpha = \int \d^2r \, \d \tau \, \psid H_\alpha \psi$, from which an expansion order by order in $\alpha$ can be carried out. The two-point function
~\eqref{eqS:St_expansion} is non-diagonal in momentum space, since $S_\alpha$ reduces the continuous translational symmetry to the discrete translational symmetry over the reciprocal lattice $\mathcal{R}$, which is the $\mathbb{Z}$-module generated by the (linearly dependent) family of vectors $\{\bm{q}_j, j=1,2,3\}$. For every vector $\bm{b}$ in $\mathcal{R}$, we define the component $G_0'(\bm{b},\bm{k},\Omega)$ of the two-point function such that
\begin{equation}\label{eqS:non_local_prop} 
\langle \mathcal{T}  \psi_{\bm{k},\Omega} \psid_{\bm{k}+\bm{q},\Omega} \rangle_0' =  \sum_{\bm{b} \in \mathcal{R}} G_0'(\bm{b},\bm{k},\Omega)\delta(\bm{b}-\bm{q}),
\end{equation}
for all momenta $\bm{k}$, $\bm{q}$ and frequency $\Omega$. We focus on how interlayer hoppings renormalize the dispersion relation, so that we are mainly interested in the translational invariant part $G_0'(\bm{k},\Omega)=G_0'(\bm{0},\bm{k},\Omega)$ of the fermionic propagator, represented in Fig.~\ref{figS:correc_propag} 
as a double line. Successive interlayer hoppings that transfer the momenta $(\eta_1 \bm{q}_{j_1}, ..., \eta_m \bm{q}_{j_m})$ in this precise order -- where $\eta_1,...,\eta_m=\pm$, with the plus sign for a hopping to the top layer, and a minus sign to the bottom layer -- give a non-zero contribution to $G_0'(\bm{k},\Omega)$ if the following conditions are met. 

(i) Total momentum is conserved, i.e. $\sum_{r=1}^{m} \eta_r \bm{q}_{j_r} = \bm{0}$. 

(ii) Consecutive hopping processes affect different layers, {\it i.e.} $\eta_{2r} = -\eta_{2r-1}$ for all  $r=1,...,n/2$.

In particular, condition (ii) forbids odd numbers of insertions, so that all correlation functions can be expanded in~$\alpha^2$, and entails that a hopping sequence is determined by the momenta and the sign of only the first hopping process $\eta = \eta_1$. Joined with condition (i), it also yields that the transfer of a momentum at one point of the diagram must be followed by the transfer of the opposite momentum at another point. Thus we can join insertions of opposite momenta by a wavy line like in Figs.~\ref{figS:self_2}-\ref{figS:self_6row}. 

We now introduce the translational part $\Sigma_\alpha(\bm{k},\Omega)$ of the self-energy as 
\begin{equation}
G_0'(\bm{k},\Omega)^{-1} = G_0(\bm{k},\Omega)^{-1} - \Sigma_\alpha(\bm{k},\Omega).
\end{equation}
The contributions to $\Sigma_\alpha(\bm{k},\Omega)$ come from the connected two-point diagrams that conserve total momentum and that cannot be cut by one stroke into two subdiagrams that conserve themselves total momentum. Expanding Eq.~\eqref{eqS:St_expansion} to sixth order in $\alpha$, we can decompose it as $\Sigma_\alpha(\bm{k},\Omega) = \Sigma_\alpha^{2}(\bm{k},\Omega) + \Sigma_\alpha^{4}(\bm{k},\Omega) +\Sigma_\alpha^{6,\text{nes}}(\bm{k},\Omega) + \Sigma_\alpha^{6,\text{row}}(\bm{k},\Omega) + \Sigma_\alpha^{6,\text{cro}}(\bm{k},\Omega)$. In the following we use the shortcut $\bar{\eta} = - \eta$ with $\eta = \pm$, and $j,l,k =1,2,3$. 
The second order contribution (Fig.~\ref{figS:self_2}) reads
\begin{equation}\label{eqS:self_hop_2} 
\Sigma_\alpha^2(\bm{k},\Omega) = \alpha^2 \sum_{\eta, j}  {T_{j}^{\bar{\eta}}} G_{0}(\k+\eta \bm{q}_j, \Omega) T_{j}^\eta.
\end{equation}
The fourth order contribution (Fig.~\ref{figS:self_4}) reads
\begin{widetext}
~\vspace{-0.4cm}
\begin{equation}\label{eqS:self_hop_4} 
\Sigma_\alpha^4(\bm{k},\Omega) = \alpha^4 \sum_{\eta, j \neq l} T_{j}^{\bar{\eta}} G_{0}(\k+\eta \bm{q}_j, \Omega) T_{l}^{\eta} G_{0}(\k+\eta \bm{q}_j - \eta \bm{q}_l, \Omega) T_{l}^{\bar{\eta}} G_{0}(\k+\eta \bm{q}_j, \Omega) T_{j}^\eta.
\end{equation}
\begin{figure*}[t!]
\centering
\subfigure{\label{figS:self_2}}\subfigure{\label{figS:self_4}}\subfigure{\label{figS:self_6nes}}\subfigure{\label{figS:self_6row}}\subfigure{\label{figS:self_6cro}}\subfigure{\label{figS:correc_propag}}
\setcounter{subfigure}{0}
\includegraphics[scale=1]{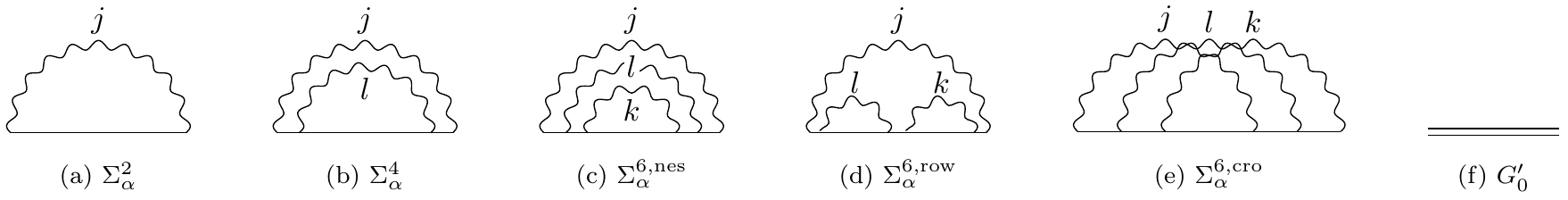}
\caption{Translational invariant part of the self-energy, $\Sigma_\alpha(\bm{k},\Omega)$, at order (a) $\alpha^2$, (b) $\alpha^4$, and (c-e) $\alpha^6$. At order six, the hopping (wavy) lines can be (c) nested, (d) in a row, or (e) crossed. Interlayer hoppings are summed over up to three momenta denoted generically as $\bm{q}_j$, $\bm{q}_l$ and $\bm{q}_k$, with $j,l,k=1,2,3$. The straight solid lines represent the fermionic propagator of the decoupled bilayer, $G_0(\k,\Omega)$, given by Eq.~\eqref{eqS:free_propag}. (f) Translational invariant part of the fermionic propagator corrected by interlayer hopping, $G_0'(\bm{k},\Omega)$, given by Eq.~\eqref{eqS:propag_corrected}.
}
\end{figure*}
The sixth order contribution splits into three terms. The first diagram hosts three nested hopping lines (Fig.~\ref{figS:self_6nes}),
\begin{multline}\label{eqS:self_hop_6nes}
\Sigma_\alpha^{6,\text{nes}}(\bm{k},\Omega) = \alpha^6 \sum_{\eta, l \neq (j,k)} T_{j}^{\bar{\eta}} G_0(\k+\eta  \bm{q}_j,\Omega) T_l^{\eta} G_0(\k+\eta  \bm{q}_j-\eta  \bm{q}_l,\Omega) T_{k}^{\bar{\eta}} G_0(\k+\eta  \bm{q}_j+\eta  \bm{q}_k-\eta  \bm{q}_l,\Omega) \cdot \\ T_{k}^{\eta} G_0(\k+\eta  \bm{q}_j-\eta \bm{q}_l,\Omega) T_l^{\bar{\eta}} G_0(\k+\eta  \bm{q}_j,\Omega) T_{j}^{\eta}.
\end{multline}
The second diagram hosts two hopping lines in a row, embedded in a third one (Fig.~\ref{figS:self_6row}),
\begin{multline}\label{eqS:self_hop_6row}
\Sigma_\alpha^{6,\text{row}}(\bm{k},\Omega) = \alpha^6 \sum_{\eta, j \neq (l,k)}  T_j^{ \bar{\eta }} G_0(\k+\eta \bm{q}_j,\Omega) T_l^{\eta} G_0(\k+\eta \bm{q}_j-\eta \bm{q}_l,\Omega) T_l^{\bar{\eta}} G_0(\k+\eta  \bm{q}_j,\Omega) \cdot \\ T_k^{\eta } G_0(\k+\eta \bm{q}_j-\eta  \bm{q}_k,\Omega) T_k^{ \bar{\eta }} G_0(\k+\eta  \bm{q}_j,\Omega) T_j^{\eta}.
\end{multline}
The third diagram consists in three crossing hopping lines (Fig.~\ref{figS:self_6cro}),
\begin{multline}\label{eqS:self_hop_6cro}
\Sigma_\alpha^{6,\text{cro}}(\bm{k},\Omega) = \alpha^6 \sum_{\eta, j \neq l \neq k} T_k^{\bar{\eta}} G_0(\k+\eta \bm{q}_k,\Omega) T_l^{\eta} G_0(\k+\eta  \bm{q}_k-\eta \bm{q}_l,\Omega) T_j^{\bar{\eta}} G_0(\k+\eta \bm{q}_j + \eta  \bm{q}_k-\eta \bm{q}_l,\Omega) \cdot \\ T_k^{\eta} G_0(\k+\eta \bm{q}_j-\eta  \bm{q}_l,\Omega) T_l^{\bar{\eta}} G_0(\k+\eta \bm{q}_j,\Omega) T_j^{\eta}.
\end{multline}
Within a low-energy theory where $k,\Omega \ll 1$, we can further expand to order two in momentum~$\bm{k}$, and one in Matsubara frequency $\Omega$, which results in
\begin{multline}
\label{eqS:sigma_four} 
\Sigma_\alpha(\bm{k},\Omega)  =  \left[3\alpha^2 -\alpha^4 (1-\beta^2)^2 + \dfrac{3\alpha^6}{49}\left(37-112 \beta ^2 +119 \beta ^4 -70 \beta ^6\right) \right] \bm{\sigma} \! \cdot \! \bm{k} \, \tau_0 \\
+ [3\alpha^2 \beta^2 - 9 \alpha^4 \beta^2 (1-\beta^2)] \begin{pmatrix}
 0 & i\underline{k}^2  \\
 \, -i{\underline{k}^*}^2 & 0
\end{pmatrix} \! \tau_z \\
+\left[3\alpha^2(1 + \beta^2) + 2\alpha^4(1+7\beta^2+4\beta^4) +  \dfrac{3\alpha^6}{28}\left(8+16\beta ^2 +376 \beta ^4 + 187 \beta ^6 \right) \right] i\Omega \sigma_0 \tau_0,
\end{multline}
where $\underline{k} = k_x + i k_y$. If one keeps only the correction to the linear dispersion, the translational part of the fermionic propagator corrected by interlayer hoppings can be massaged into 
\begin{equation}
\label{eqS:propag_corrected} 
G_0'(\bm{k},\Omega) = N_\psi^{-1} (v\bm{\sigma} \! \cdot \! \bm{k}-i\Omega)^{-1}
\end{equation}
with the normalisation of the wave function $N_\psi = 1+3\alpha^2(1 + \beta^2) + 2\alpha^4(1+7\beta^2+4\beta^4) + \frac{3}{28}\alpha^6(8+16\beta ^2 +376 \beta ^4 + 187 \beta ^6)$, and the Fermi velocity dressed by interlayer hoppings
\begin{equation}\label{eqS:fermi_velocity} 
v=\dfrac{1-3\alpha^2+\alpha^4 \left(1-\beta^2\right)^2 - \frac{3}{49}\alpha^6 \left(37-112 \beta ^2 +119 \beta ^4 -70 \beta ^6\right)}{1+3\alpha^2(1 + \beta^2) + 2\alpha^4(1+7\beta^2+4\beta^4) + \frac{3}{28}\alpha^6(8+16\beta ^2 +376 \beta ^4 + 187 \beta^6)},
\end{equation}
as given in Eq.~\eqref{eq:fermi_velocity}. We remind that $v$ is expressed in units of the Fermi velocity $v_0$ of monolayer graphene.
\end{widetext}

\section{Symmetries of the model}
\setcounter{equation}{0}
\label{app:symmetries}

\subsection{Complete symmetries}

\begin{table*}[t!]
\renewcommand{\arraystretch}{1.5}
\setlength\tabcolsep{0.36cm}
\begin{tabularx}{\textwidth-0\columnsep}{lccccccc}
\hline
\hline
 Class & $e$ & $\bar{e}$ & $2C_3$ & $2\bar{C_3}$ & $2PC_2$ & $2PC_2C_3$ & $2 P C_2C_3^2$\\
\hline
 Elements & $\{e\}$ & $\{\bar{e}\}$ & $\{C_3,C_3^2\}$ & $\{\bar{C_3}, \bar{C_3^2}\}$ & $\{PC_2, \bar{P}C_2\}$ & $\{P C_2 C_3, \bar{P} C_2 C_3^2\}$ & $\{\bar{P} C_2 C_3, P C_2 C_3^2\}$\\
\hline
\hline
Class &  \multicolumn{3}{c}{$6P$} & \multicolumn{3}{c}{$6C_2$} &\\
Elements & \multicolumn{3}{c}{$\{P, P C_3, P C_3^2, \bar{P}, \bar{P} C_3, \bar{P} C_3^2\}$} & \multicolumn{3}{c}{$\{C_2, C_2 C_3, C_2 C_3^2, \bar{C}_2, \bar{C}_2 C_3, \bar{C}_2 C_3^2\}$} & \\
\hline
\hline
\end{tabularx}
\caption{Classes of conjugation of the unitary group $\tilde{D}_3$, with their names (first line) and their elements (second line). $e$ is the identity operator, $\bar{e}$ the $2\pi$ rotation of a spin one half, and $\bar{R}$ denotes the product $\bar{e}R$ for any operator $R$.}
\label{tab:classes_conjug}
\end{table*}

The symmetries of the single-particle Hamiltonian  $H_0' = H_0+H_\alpha$ of Eq.~\eqref{eq:single_hamilt} are highly constrained by the interlayer hopping term. Indeed, the Hamiltonian of the decoupled bilayer $H_0= i\bm{\sigma} \! \cdot \! \bm{\partial} \tau_0$,
is invariant under the layer pseudospin rotational group U$(1)$, and the Poincar\'e group $\mathbb{R}^{1+2} \rtimes \text{O}(1,2)$, where $\rtimes$ indicates a semi-direct product. The hopping Hamiltonian $H_\alpha=\alpha \sum_{j=1}^3 e^{-i \bm{q}_j\cdot\bm{r}} T_j^+ + \text{h.c}$,
breaks Lorentz invariance, continuous space translations and layer plus pseudospin rotational symmetry. The symmetry group is thus reduced to the symmorphic space-time group $\mathbb{R}  \times (\bm{a_1} \mathbb{Z} + \bm{a_2} \mathbb{Z}) \rtimes D_3$, composed of time translation $\mathbb{R}$, discrete translations on the moir\'e lattice $\bm{a_1} \mathbb{Z} + \bm{a_2} \mathbb{Z}$, where $\bm{a_1}$ and $\bm{a_2}$ are the superlattice vectors, and the point group $D_3$ generated by the rotation $C_3$ around the $z$ axis and the rotation $C_2$ around the $x$ axis,
\begin{equation}
D_3 = \{e,C_3,C_3^2,C_2,C_2C_3,C_2C_3^2\}.
\end{equation}

Henceforth we disregard the translational symmetries and focus on the magnetic group generated by the point group $D_3$ and the two special ``symmetries" $IT$ and $P$. These operations act by conjugation on the single-particle Hamiltonian $H_0'$ in a four-dimensional representation ($4$d rep.), denoted as $\Gamma$, whose unitary matrix representation we now define.

The operation $C_3$ rotates the bilayer by an angle $2\pi/3$ around the $z$ axis perpendicular to the bilayer
. Only the sublattice pseudospin is rotated, while the layer pseudospin is unaffected, so that
\begin{equation}
\label{eqS:C3_Grep}
\Gamma(C_3) = e^{2i\pi/3 \sigma_z} \tau_0.
\end{equation}
The operation $C_2$ rotates the bilayer by an angle $\pi$ around the $x$ axis of Fig.~\ref{fig:BZ_mBZ} at mid-distance between the layers. Both sublattices and layers are flipped, so that
\begin{equation}
\Gamma(C_2) = \sigma_x \tau_x.
\end{equation}
The composition of inversion $I$ and time reversal $T$, denoted as $IT$, is antiunitary and represented by
\begin{equation}
\label{eqS:IT_Grep}
\Gamma(IT) = \sigma_x \tau_0 \mathcal{K},
\end{equation}
where $\mathcal{K}$ denotes the complex conjugation of the matrix elements. The operations $R$ defined in Eq.~\eqref{eqS:C3_Grep} to \eqref{eqS:IT_Grep} are pure symmetries of the Hamiltonian, which means that $\Gamma(R)^{-1}H_0'(R\bm{r},Rt)\,\Gamma(R) = H_0'(\bm{r},t)$ for $R=C_3,C_2$, and $\Gamma(R)^{-1}H_0'(R\bm{r},Rt)^*\,\Gamma(R) = H_0'(\bm{r},t)$
for the antiunitary element $R=IT$. Finally, the unitary particle-hole operation $P$ reverses the energy, and acts in real space as a reflection $x \mapsto -x$. Its matrix representation reads
\begin{equation}
\label{eqS:C_Grep}
\Gamma(P) = \sigma_x \tau_z.
\end{equation}
Following Ref.~\onlinecite{hejazi_multiple_2019} we define $P$ as a unitary operation in order to have a single antiunitary generator ($IT$), unlike the convention of Ref.~\onlinecite{song_all_2018}.
This operation is an antisymmetry of the Hamiltonian, which means that $\Gamma(P)^{-1}H_0'(P\bm{r},Pt)\,\Gamma(P) = -H_0'(\bm{r},t)$. This antisymmetry is lost when the angular dependence of the kinetic terms $\bm{\sigma}_{\pm\theta/2} \cdot \bm{k}$ is kept, when terms quadratic in momentum are included in the single-particle Hamiltonian, or when intervalley scattering is permitted~\cite{song_all_2018}. Since $\Gamma$ is faithful rep., we can infer the multiplication table of the magnetic group from that of the matrix representation of the four generators.

\subsection{Unitary group}

The unitary group $\tilde{D}_3$ generated by $D_3$ and $P$ can be cast into the semi-direct product
\begin{equation}
\tilde{D}_3 = \{e,P,\bar{e},\bar{P}\} \rtimes D_3,
\end{equation}
where $\bar{e}=(PC_2)^2$ and a barred operator represents the product of this operator by $\bar{e}$. The classes of conjugation of $\tilde{D}_3$ are given in Tab.~\ref{tab:classes_conjug}. The double group $D_3^D = D_3 \times \{e,\bar{e}\}$ is a normal subgroup of $\tilde{D}_3$, while $D_3$  is not. To find the irreducible representatinos (irrep.) of $\tilde{D}_3$---listed in Tab.~\ref{tab:characters}---we can either find them from scratch using the composite operator method~\cite{chen2002group}, or construct them by induction and other tricks from that of $D_3$. Let us illustrate the second method.

\begin{table*}[t!]
\begin{center}
\renewcommand{\arraystretch}{1.3}
\setlength\tabcolsep{0.55cm}
\begin{tabularx}{\textwidth-0\columnsep}{cccccccccc}
\hline
\hline
~Irrep. & $e$ & $\bar{e}$ & $2C_3$ & $2\bar{C}_3$ & $2PC_2$ & $2PC_2C_3$ & $2PC_2C_3^2$ & $6P$ & $6C_2$\\
\hline
~$A_1$ & $1$ & $1$ & $1$ & $1$ & $1$ & $1$ & $1$ & $1$ & $1$\\
~$A_2$ & $1$ & $1$ & $1$ & $1$ & $-1$ & $-1$ & $-1$ & $1$ & $-1$\\
~$a_1$ & $1$ & $1$ & $1$ & $1$ & $-1$ & $-1$ & $-1$ & $-1$ & $1$ \\
~$a_2$ & $1$ & $1$ & $1$ & $1$ &  $1$ & $1$ & $1$ & $-1$ & $-1$\\
~$E_1$ & $2$ & $-2$ & $-1$ & $1$ & $0$ & $\sqrt{3}$ & $-\sqrt{3}$ & $0$ & $0$\\
~$E_2$ & $2$ & $2$ & $-1$ & $-1$ & $2$ & $-1$ & $-1$ & $0$ & $0$\\
~$E_3$ & $2$ & $-2$ & $2$ & $-2$ & $0$ & $0$ & $0$ & $0$ & $0$\\
~$E_4$ & $2$ & $2$ & $-1$ & $-1$ & $-2$ & $1$ & $1$ & $0$ & $0$\\
~$E_5$ & $2$ & $-2$ & $-1$ & $1$ & $0$ & $-\sqrt{3}$ & $\sqrt{3}$ & $0$ & $0$\\
\hline
\hline
\end{tabularx}
\end{center}
\caption{Table of characters of the unitary group $\tilde{D}_3$. Each column corresponds to a class of conjugation, and each line to an irrep. We use the symbols $A$ and $E$ prescribed by Mulliken's notation; the symbol $a$ denotes a $1$d irrep. whose character differs than one on antisymmetric operators.}
\label{tab:characters}
\end{table*}

The two $1$d irrep. of the quotient group $\tilde{D}_3/D^D_3 = \{e,P\}$ generate the irreps $A_1$ and $a_1$. The irrep. $A_1$ and $A_2$ of $D_3$ induce the reps $A_1\! \uparrow \!\tilde{D}_3 \sim A_1 \oplus a_1 \oplus E_3$ and $A_2\! \uparrow \!\tilde{D}_3 \sim A_2 \oplus a_2 \oplus E_3$ respectively, where $\oplus$ denotes a direct sum and $\sim$ the equivalence of rep. The commutator subgroup $[\tilde{D}_3,\tilde{D}_3]$ is isomorphic to $D_3$, whose index $|\tilde{D}_3/D_3|=4$ gives the number of $1$d irrep. Hence we have found all $1$d irrep. The cardinal of the group being $|\tilde{D}_3|=24$, the remaining irrep. are five $2$d irrep., including $E_3$. We can decompose the $4$d rep. $\Gamma$ defined by Eq.~\eqref{eqS:C3_Grep} to \eqref{eqS:IT_Grep} and \eqref{eqS:C_Grep} as $\Gamma \sim E_1 \oplus E_5$. The $2$d irrep. $E$ of $D_3$ induces the rep. $E\! \uparrow \!\tilde{D}_3 \sim E_1 \oplus E_2 \oplus E_4 \oplus E_5$, where the remaining irrep. $E_2$ and $E_4$ can be found by orthonormality of the characters.

\subsection{Magnetic group}
The group generated by $\tilde{D}_3$ and $IT$ is the dichromatic magnetic group
\begin{equation}
\mathcal{M} = \tilde{D}_3 \times \{e,IT\}.
\end{equation}
To find the corepresentations (corep.) of $\mathcal{M}$, we apply the Schur-Frobenius criterion~\cite{Ma:2007,Atkins:1970,Dresselhaus:2007,Woit:2017,Zhong:2004} to each irrep. of $\tilde{D}_3$. The Schur-Frobenius criterion states the following. For any irrep $\rho$ of $\tilde{D}_3$, let us define the rep. $\rho' : \, \tilde{D}_3 \rightarrow M_n(\mathbb{C}), R \mapsto \rho(IT \!\cdot\! R\! \cdot \! IT^{-1})^* $.
We are necessarily in one of the three following scenarios. (i,ii) Either $\rho$ is equivalent to its primed counterpart, in which case there exists an invertible matrix $U$ such that $\rho' = U \rho U^{-1}$. (i) if $UU^* = \rho(IT^2)$, there is no Kramer degeneracy: the corep. issued from $\rho$ has the same dimension as $\rho$ and satisfies $\rho(IT)=\pm U$. (ii) If $UU^* = -\rho(IT^2)$, the corep. issued from $\rho$ has twice the dimension. (iii) Or $\rho$ is not equivalent to its primed counterpart, in which case $\rho'$ is necessarily equivalent to another irrep. of $\tilde{D}_3$, and the corep. has again twice the dimension of $\rho$, and coincides with $\rho \oplus \rho'$ on $\tilde{D}_3$.

It turns out that all irrep. of $\tilde{D}_3$ pertain to case (i), except $E_1$ and $E_5=E_1'$, which fall into case (iii). In the former case, each irrep. leads to two corep., with $\rho(IT) = \pm 1$ for the $1$d irrep. or $\rho(IT) = \pm \sigma_x$ for the $2$d irrep., where $\sigma_x$ represents here a generic Pauli matrix, but has nothing to do with the pseudospin. In the latter case, the one corep. formed by $E_1$ and $E_5$ is equivalent to the 4d rep. $\Gamma$. In the following, we write each corep. with an exponent $\pm$ to indicate whether the eigenvalues of $\rho(IT)$ are $+1$ or $-1$.

\subsection{Quartic interactions}

The direct product $\Gamma^\dagger \otimes \Gamma$ dictates how the bilinear $\psid_i \psi_j$ with $i,j=1, ...,4$, transforms under the magnetic group~$\mathcal{M}$. The Clebsch-Gordan series reads
\begin{eqnarray}
X(\Gamma^\dagger \otimes \Gamma)X^{-1} &=& A_1^ +  \oplus a_1^+  \oplus A_2^+  \oplus a_2^+  \oplus A_1^-   \oplus a_1^- \nonumber \\
&& \!\!\!\!\!\!\!  \!\!\!\!\!\!\!  \oplus A_2^-  \oplus a_2^- 
 \oplus E_2^+  \oplus E_4^+  \oplus E_2^-  \oplus  E_4^-, \ \ \ \ \ \  \
\label{eqS:clebsch_series}
\end{eqnarray}
where the transformation matrix $X$ contains the Clebsch-Gordan coefficients, which can be found using the formula~\cite{Ma:2007}
\begin{equation}
X_{ik,\gamma m} X^*_{jl,\gamma n}= \dfrac{n_\gamma}{|\mathcal{M}|}\sum_{R \in \mathcal{M}} \rho_\gamma(R)^*_{mn} \Gamma^\dagger(R)_{ij} \Gamma(R)_{kl},
\end{equation}
where $\rho_\gamma$ is the $\gamma^{\rm th}$ irrep. in the series \eqref{eqS:clebsch_series}, with dimension $n_\gamma$, and $|\mathcal{M}|=48$ is the cardinal of the magnetic group. The coefficients of the matrices $\bm{M_\gamma}$ of Eq.~\eqref{eq:inter_action} that transforms by conjugation according to the irrep. $\rho_\gamma$ are listed in the $\gamma^{\rm th}$ column of $X$, i.e. for $a,b=1, ..., 4$, the two components of the vector $\bm{M_\gamma}$ read
\begin{align}
\label{eqS:matrix_clebsch}
 (M_{\gamma}^{(1)})_{ab} = X_{ab,\gamma 1}, & &  (M_{\gamma}^{(2)})_{ab} = X_{ab,\gamma 2}.
 \end{align}
To find the quartic interaction that preserve the magnetic group $\mathcal{M}$ we must find all copies of the trivial irrep. $A_1^+$ into the product $(\Gamma^\dagger \otimes \Gamma) \otimes (\Gamma^\dagger \otimes \Gamma)$. By inspecting the characters, it is clear that only products of the same irrep. decompose themselves into a copy of $A_1^+$. For the $1$d irrep. $\rho = A_1^+, a_1^+, A_2^+,a_2^+,A_1^-,a_1^-,A_2^-$ and $a_2^-$, the decomposition is simply
\begin{equation}
\rho \otimes \rho = A_1^+.
\end{equation}
Thus the quartic interaction corresponding to one of these irreps is of the form $M \otimes M$, where $M$ is found applying Eqs.~\eqref{eqS:clebsch_series} and \eqref{eqS:matrix_clebsch}. The interaction matrices for these eight $1$d irreps are listed in Tab.~\ref{tab:matrix_rep}.
For the $2$d irreps, we have
\begin{eqnarray}
E_2^\eta \otimes E_2^\eta \sim A_1^\eta \oplus A_2^\eta \oplus E_4^\eta, \\
E_4^\eta \otimes E_4^\eta \sim A_1^\eta \oplus A_2^\eta \oplus E_2^\eta,
\end{eqnarray}
for $\eta=\pm$. For each of these irreps, the invariant combination transforming as $A_1^+$ is $M \otimes N$, where $(M, N)$ is the basis of the two-dimensional space on which the irrep acts. The interaction matrices for these four $2$d irreps are listed in Tab.~\ref{tab:matrix_rep}.

\section{\label{appendix-mean-field} Mean-field theory}
Here we consider only the four relevant instabilities related to interactions proportional to  $\sigma_z \tau_{0/z}$ and $\bm{\sigma} \tau_{0/z}$ with coupling constants $g_{0/z}$ and $\lambda_{0/z}$ respectively. The corresponding order parameters, which we denote $\Delta_{0/z}$ and $\bm{\mathcal{G}}_{0/z}$, can be found in the mean-field approximation by solving the appropriate self-consistent equations.  
For the sake of simplicity we write down these equations separately
for each instability. Upon introducing an ultraviolet cut-off $\Lambda$, these equations read
\begin{subequations}
\begin{align}
\Delta_{0/z} & = -2 g_{0/z}  
\displaystyle\int \!
\mathrm{d} \omega \int_\Lambda \dfrac{
\mathrm{d}^{2} q}{(2\pi)^3} \langle \psid_{q,\omega} \sigma_z \tau_{0/z}
\psi_{q,\omega}\rangle, \\
\bm{\mathcal{G}}_{0/z} & = -2 \lambda_{0/z} 
\displaystyle\int \!
\mathrm{d} \omega \int_\Lambda \dfrac{
\mathrm{d}^{2} q}{(2\pi)^3} \langle \psid_{q,\omega} \bm{\sigma} \tau_{0/z}
\psi_{q,\omega}\rangle,
\end{align}\label{eqS:MFselfconsistent1}
\end{subequations}
$\!\!\!$where the momentum integral runs over a square of side $\Lambda$.  The correlators in Eq.~\eqref{eqS:MFselfconsistent1}
 are the translationally-invariant parts of statistical averages computed over the Bloch Hamiltonian density 
$H_{\rm MF}' = H_0' 
+ \bm{\sigma} \cdot (\bm{\mathcal{G}}_0 \tau_0 + \bm{\mathcal{G}}_z \tau_z) +  \sigma_z (\Delta_0 \tau_0 +\Delta_z \tau_z )$. 
The perturbative expansion of the correlators in $\alpha$ 
can be done along the lines of computing the self-energy. We have previously found that the propagator corrected by interlayer hoppings $G_0'$ is of the form \eqref{eqS:propag_corrected}; similarly, the mean-field Hamiltonian becomes
\begin{multline}
    H_{\rm MF}' \mapsto N_\psi \left[ v \bm{\sigma} \! \cdot \! \left(\left(\bm{k} +N^{(\mathcal{G})}_{0}\bm{\mathcal{G}}_0 \right)\tau_0 + N^{(\mathcal{G})}_{z} \bm{\mathcal{G}}_z \tau_z \right) \right. \\ \left.  + \sigma_z \left(N^{(\Delta)}_{0} \Delta_0 \tau_0 + N^{(\Delta)}_{z} \Delta_z \tau_z \right) \right]. \qquad
\end{multline}
The effect of interlayer hoppings is to enhance the order parameters 
by factors $N^{(G/\Delta)}_{0/z}(\alpha,\beta)$, which are the counterparts of the renormalized Fermi velocity for 
the matrix structures corresponding to those order parameters. They are calculated diagrammatically to sixth order in $\alpha$ and satisfy
\begin{widetext}
\begin{subequations}
\begin{align}[left = \empheqlbrace\,]
\label{eq:u_w} 
N_{\psi} N_{0}^{(\Delta)} &= 1+3\alpha^2 \left(1-\beta^2\right) + 2\alpha^4 \left(1-\beta^2\right) \left(1+2\beta^2\right) + \frac{1}{28} \alpha^6 \left(24 - 80 \beta ^2 + 352 \beta ^4 - 233 \beta ^6\right),\\
N_{\psi} N_{z}^{(\Delta)} &= 1-3\alpha^2 \left(1-\beta^2\right) + 2\alpha^4 \left(1-\beta^2\right) \left(1-4\beta^2\right) - \frac{1}{28} \alpha^6 \left(56 - 304 \beta ^2 + 872 \beta ^4 -561 \beta ^6\right),\\
v N_{\psi} N_{0}^{(\mathcal{G})} &= 1-3\alpha^2+\alpha^4 \left(1-\beta^2\right)^2 - \frac{3}{49}\alpha^6 \left(37-112 \beta ^2 +119 \beta ^4 -70 \beta ^6\right),\\
v N_{\psi} N_{z}^{(\mathcal{G})} &= 1+3\alpha^2+\alpha^4 \left(1+10\beta^2+\beta^4\right) + \frac{3}{49} \alpha^6 \left(9 + 441 \beta ^4 + 70 \beta ^6\right), 
\end{align}
\end{subequations}
where the wavefunction normalization $N_{\psi}$ is given in Sec.~\ref{sec:non-int_dispersion}. In the main text, we plotted the order parameters corrected by interlayer hoppings, {\it i.e.} the quantities $\Delta_{0/z}' = N_{0/z}^{(\Delta)} \Delta_{0/z}$ and $\mathcal{G}_{0/z}' = N_{0/z}^{(\mathcal{G})} \mathcal{G}_{0/z}$, which 
satisfy the self-consistency equations
\begin{subequations}
\begin{align}
\Delta_{0/z}' & = \dfrac{g_{0/z} N_{0/z}^{(\Delta)} \Lambda^2}{v N_\psi}
F\left(\Delta_{0/z}'\right), \\
\mathcal{G}'_{0/z} & = \dfrac{\lambda_{0/z} N_{0/z}^{(\Delta)} \Lambda^2}{N_\psi}
F_{0/z}\left(\mathcal{G}'_{0/z}\right),
\label{eqS:MFselfconsistent}
\end{align}
\end{subequations}
where for simplicity we assume the shift momenta to be aligned along a
crystallographic axis of the moire pattern, here along the $y$ axis. The dimensionless functions $F$ and $F_{0/z}$ read
\begin{subequations}
\begin{align}
F(x) & = \frac{2 x}{\pi^2} \left[-\log \left(\sqrt{x^2+2}-1\right)+\log
   \left(\sqrt{x^2+2}+1\right)-2 x \cot ^{-1}\left(x
   \sqrt{x^2+2}\right)+2 \coth
   ^{-1}\left(\sqrt{x^2+2}\right)\right], \\
F_0(x) & = \frac{1}{\pi ^2} \left[-\sqrt{y_-}+\sqrt{y_+}+\tanh
   ^{-1}\left(\sqrt{y_-}\right)-\tanh
   ^{-1}\left(\sqrt{y_+}\right)-y_- \coth
   ^{-1}\left(\sqrt{y_-}\right)+y_+ \coth
   ^{-1}\left(\sqrt{y_+}\right)\right], \\
F_z(x) & = \frac{1}{2\pi^2} \left[ x^2 \log \left(\dfrac{1-x}{1+x}\right)-2x^2 \tanh ^{-1}(x)+\left(1-2x\right) \log
   \left(\sqrt{y_-}-1\right)-\left(1+2x\right) \log
   \left(\sqrt{y_+}-1\right) +
   \right. \nonumber \\ &  z_+ \log
   \left(\sqrt{y_+}+1\right) -z_- \log
   \left(\sqrt{y_-}+1\right) + 2\left(\sqrt{y_+}-\sqrt{y_-}\right) \bigg],
\label{eqS:MFselfconsistent2}
\end{align}
\end{subequations}
where $y_\pm = 2 + x(x\pm 2)$ and $z_\pm = 1 + 2x(x\pm 1)$.
\end{widetext}

\section{\label{sec:diagrams_int} Renormalization  } 
\setcounter{equation}{0}
\subsection{Hubbard-Stratonovitch decoupling}
\label{sec:hubbard_decoupl} 

We aim at finding the most relevant insulating state near charge
neutrality. It is therefore practical to decouple the interactions in
the direct, particle-hole channel, to evince order parameters of the
form $\langle \psid M \psi \rangle$ for $M \in \{R_i,\bm{M}_j\}$,
where the bracket $\langle ... \rangle$ denotes the ensemble average
over the complete action $S = S_0' + S_{\rm int}$.

Using Hubbard-Stratonovitch transformations, we introduce one auxiliary bosonic field for each interaction, whose ground state value in the correlated phase is a constant solution of the classical equation of motion. We must distinguish between the $1$d corep., for which a scalar field $\phi_i$ for $i=1,...,8$, is sufficient, and the $2$d corep., for which a two-component field $\bm{\varphi}_j = \{\varphi_{j,1},\varphi_{j,2}\}$ must be introduced, for $j=9,...,12$. Such transformation enables to recast the action for quartic fermion interactions~\eqref{eq:inter_action} into
\begin{eqnarray}
&& \!\!\!\!\!\!\!\!\!\!\!\!  S_{\rm int}[\psi^\dagger,\psi]  \rightarrow S_{\text{Hub}}[\psi^\dagger,\psi,\phi] \nonumber\\ 
&&= \sum_{i=1}^{8} \int \d^2r \, \d \tau \, \left(\phi_i^2 + 2\sqrt{g_{i}}\,\psid \phi_i R_i \psi \right) \nonumber\\ 
&&+ \sum_{j=9}^{12} \int \d^2r \, \d \tau \, \left(\bm{\varphi}_j^2 + 2\sqrt{\lambda_{i}}\,\psid \bm{\varphi}_j \cdot \bm{M}_j \psi \right),
\label{eqS:Hubbard_Strato} 
\end{eqnarray}
where the sum runs over both $1$d and $2$d irreps.. For simplicity we dropped the spatial and time dependences of the fields in Eq.~\eqref{eqS:Hubbard_Strato}. The action for quartic fermion interactions thus splits into a bosonic quadratic action (the first term in the parenthesis), and a three-point vertex which takes the form of a Yukawa coupling (the second term in the parenthesis). 

\subsection{Renormalization procedure}
\label{sec:renorm_proc}

The field theory described by the sum of action $S_0'$ and action $S_{\text{Hub}}$~\eqref{eqS:Hubbard_Strato} has critical dimension $d_c=2$, which entails that in $d=3$ space-time dimensions, we expect that all interactions lead to quantum critical points that are perturbative in the small parameter $\epsilon=d-2$. 
From now onwards, we work in Fourier space and express all fields and
integrals in terms of the momentum $\bm{q}$ and the Matsubara
frequency $\omega$. A general Fourier-transformed field is written
$\phi_{\bm{q},\omega}$ for the bosonic case, or $\{
\psi_{\bm{q},\omega},\psid_{\bm{q},\omega} \}$ for the fermionic case,
where $\psid_{\bm{q},\omega} = (\psi_{\bm{q},\omega})^{\dagger}$
denotes the conjugate of the Fourier-transformed field
$\psi_{\bm{q},\omega}$. 
To renormalize the field theory, we assume that the complete action $S$ is actually expressed in terms of bare 
fields $\{\mathring{\phi},\mathring{\psi}\}$ for $\phi \in \{\phi_i,\bm{\varphi}_j\}$ and couplings $\mathring{g}$ for $g \in \{g_i,\lambda_j\}$ , which are ill-defined in the interacting theory. The physical parameters and fields---written without the $\mathring{}$ symbol---are connected to their bare counterparts through the so-called $Z$ constants.

We define define the $Z$ constants for the fields such that
\begin{equation} \label{eqS:Z_parama}
\mathring{\phi}  = Z_{\phi}^{1/2} \phi, ~~~\mathring{\psi} =  Z_{\psi}^{1/2} \psi.
\end{equation}
To regularize the theory, we work in an isotropic space-time of
dimension $d=2+\epsilon$, and introduce a mass scale $\mu$ to make the
regularized couplings dimensionless. 
A renormalized coupling $g$ is linked to its bare value $\mathring{g}$ by
\begin{equation}
\label{eqS:Z_couplings}
 \mathring{g} = \mu^{-\epsilon} N_\psi^2 Z_{g}^2 Z_{\phi}^{-1}g.
\end{equation}
We included the normalization of the wavefunction $N_\psi$ in the redefinition of the couplings in order to compensate at all loop orders those arising from the corrected fermionic propagator $G_0'$, given in Eq.~\eqref{eqS:propag_corrected}. Owing to dimensional regularization, we must promote the Pauli matrices in $S_0$ to a Clifford algebra in arbitrary dimension $d$, satisfying the anticommutation rules $\{\sigma_i,\sigma_j\} = 2 \delta_{ij}$ for $i,j=1,...,d$.  Using Eqs.~\eqref{eqS:Z_parama} and \eqref{eqS:Z_couplings}, we find the renormalized action $S_{\rm R} = S_{\text{R},0} + S_{\text{R},\alpha} + S_{\text{R},\phi} + S_{\text{R},\text{int}}$, where the quadratic, decoupled action reads
\begin{equation}
\label{eqS:SR0} 
S_{\text{R},0} = \int_{\bm{q},\omega} \psid_{\bm{q}} ( 
\bm{\sigma}\!\cdot\!\bm{q} \tau_0 -i \omega \sigma_0 \tau_0) \psi_{\bm{q}}.
\end{equation}
The quadratic hopping action reads
\begin{equation}
S_{\text{R},\alpha} = 
\alpha \sum_{\eta} \sum_{j=1}^3 \int_{\bm{q},\omega} \psi^\dagger_{\bm{q}}T_j^\eta \psi_{\bm{q}+\eta\bm{q}_j}.
\end{equation}
The renormalized bosonic part of the  action is given by
\begin{equation}
\label{eqS:SRphi} 
S_{\text{R},\phi} = \sum_{i=1}^{8} Z_{\phi_i} \int_{\bm{q},\omega} (\phi_i^2)_{\bm{q}} + \sum_{j=8}^{12} Z_{\bm{\varphi}_j} \int_{\bm{q},\omega} (\bm{\varphi_j}^2)_{\bm{q}}.
\end{equation}
Finally, the renormalized interaction between fermionic and bosonic fields is
\begin{eqnarray}
 S_{\text{R},\rm int} &=& 2\mu^{-\epsilon/2} N_\psi \left[ \sum_{i=1}^{8} Z_{g_i} \sqrt{g_i}\int_{\bm{q}, \omega_q, \bm{p},\omega_p} \psid_{\bm{q}} (\phi_{i})_{\bm{q}-\bm{p}} R_i \psi_{\bm{p} } \right. \nonumber \\
&& \!\!\!\! \left. + \sum_{j=1}^{12} Z_{\lambda_j} \sqrt{\lambda_j}\int_{\bm{q}, \omega_q, \bm{p},\omega_p} \psid_{\bm{q}} (\bm{\varphi}_{j})_{\bm{q}-\bm{p}}\!\cdot\!\bm{M}_j \psi_{\bm{p} } \right].\ \ \ \ \ 
\label{eqS:SRint} 
\end{eqnarray}
In Eqs.~\eqref{eqS:SR0} -\eqref{eqS:SRint}, we have ommitted dependence of the fields on frequency and used the shorthand
\begin{equation}
\int_{\bm{q},\omega} = \int_{\mathbb{R}^{d}} \dfrac{\mathrm{d}^{d-1} q \, \mathrm{d} \omega}{(2\pi)^d}.
\end{equation}
To fix values of the $Z$ constants we use the minimal subtraction
(MS) scheme,  
i.e.\ we absorb in them only the divergent parts of the diagrams. Inspecting Eqs.~\eqref{eqS:SRphi} and \eqref{eqS:SRint}, we see that the constant $Z_\phi$ can be found from the divergences of the polarization, i.e.\ the bosonic self-energy, while the constant $Z_g$ can be determined by absorbing the divergences of the three-point vertices. 
Before computing explicitely the diagrams for the polarisation and vertices let us outline the general strategy.

\subsection{Preliminary mathematical remarks }

\begin{figure*}[t!]
\centering
\subfigure{\label{figS:polar_0}}\subfigure{\label{figS:polar_2}}\subfigure{\label{figS:vertex_0}}\subfigure{\label{figS:vertex_int}}\subfigure{\label{figS:vertex_ext}}\subfigure{\label{figS:vertex_iso}}\subfigure{\label{figS:vertex_cro}}
\setcounter{subfigure}{0}
\includegraphics[scale=1]{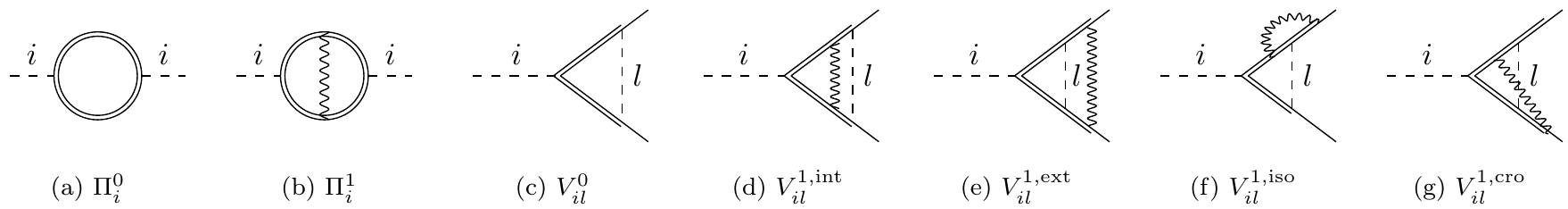}
\caption{One-particle irreducible diagrams at one loop, up to order two in interlayer hoppings. The double line stands for the fermionic propagator corrected by interlayer hoppings of Fig.~\ref{figS:correc_propag}, the dashed line for the bosonic propagator, and the wavy line for the sum of interlayer hoppings  of opposite momenta $\pm\eta \bm{q}_j$, for $\eta=\pm$ and $j=1,2,3$. Polarisation $\Pi_i$ at zero external momentum and fixed Matsubara frequency, for the field $\bm{\phi_i}$ at order (a) $\alpha^0$ and (b) $\alpha^2$. Three-point vertex $V_{il}$ at order (c) $\alpha^0$ and (d-g) $\alpha^2$, whose hopping line is (d) internal, (e) external, (f) isolated and (g) crossed. }
\label{figS:1PI_diagrams}
\end{figure*}

Taking into account the interlayer hopping is done into two steps.
We first draw the diagrams without insertion the hopping matrices,   and then replace all solid lines by double lines. This corresponds to replacing free propagators by the propagators dressed by interlayer hoppings, like in the polarization shown in Fig.~\ref{figS:polar_0}. In second step we include wavy lines, i.e. hopping matrices, connecting different propagators (see Fig.~\ref{figS:polar_2}). This splitting allows us to explicitly extract factors of $v^{-1}$, where $v$ is the Fermi velocity corrected by interlayer hoppings~\eqref{eqS:fermi_velocity} and which vanishes at the first magic angle.   We restrict our computation to order $\alpha^2$, which is the first non-trivial order.

When expanding product of matrices and integrating the trace, useful relations can be found in 
Refs.~\cite{srednicki2007quantum,
kleinert2001critical,
zinn2002quantum}. The Feynman trick,
\begin{equation}
\label{eqS:1surAB}
\dfrac{1}{AB} = \int_0^1 \dfrac{\d x}{Ax+B(1-x)},
\end{equation}
valid for any expressions $A$ and $B$, enable to linearize products of denominators. For the four relevant interactions we consider here, the space-time integrals are isotropic and can be computed in arbitrary dimension $d$ using
\begin{multline}
\label{eqS:integral_dimensionalreg}
\int \dfrac{\d^d Q}{(2\pi)^d} \dfrac{Q^{2a}}{(Q^2 + m^2)^b} \\ = \dfrac{\Gamma(b-a-d/2)\Gamma(a+d/2)}{(4\pi)^{d/2} \Gamma(b)\Gamma(d/2)} m^{-2(b-a-d/2)},
\end{multline}
for any reals $a$ and $b$, and where $Q=(\bm{q},\omega)$ is the relativistic $d$-momentum. The dummy mass $m \rightarrow 0$ plays the role of an infrared regulator and $\Gamma$ denotes Euler's Gamma function, which satisfies
\begin{equation}
\label{eqS:Gamma_property}
\Gamma(-n+x) = \dfrac{(-1)^n}{n!}\left[\dfrac{1}{x} + \Psi(n+1) + \mathcal{O}(x)\right]
\end{equation}
for all real $x$ and integer $n$; this relation is usually used with $x=\epsilon$. In Eq.~\eqref{eqS:Gamma_property}, $\Psi = (\ln \Gamma)'$ is Euler's Digamma function, which does not intervene at one loop, since we discard all finite quantities in the MS scheme.

\subsection{Polarization}

The one-loop polarisation $\Pi_i$ is the self-energy of the auxiliary field $\bm{\phi_i}$ (for $i=1, ..., 12$). If $\Delta_i(\bm{k},\Omega)$ denotes the corrected propagator of the bosonic field, we have 
$\Delta_i^{-1}(\bm{k},\Omega) = Z_{\phi} -\Pi_i(\bm{k},\Omega)$. 
The one-loop diagrams contributing to the polarization at zero external momentum $\bm{k}=\bm{0}$ and fixed frequency $\Omega$ are drawn in Fig.~\ref{figS:1PI_diagrams}. 
The polarization at order $\alpha^0$ 
reads
\begin{equation}
\Pi^{0}_i = -4 g_i N_\psi^2 \displaystyle\int_{\bm{q},\omega} \text{Tr}[\bm{M}_i G_0'(\bm{q},\omega)\!\cdot\!\bm{M}_i G_0'(\bm{q},\omega)].
\label{eqS:polari_1}
\end{equation}
Notice that for the sake of generality, we will write all interaction matrices as the vectors $\bm{M}_i$, which can either denote a single matrix $R_i$ for $i=1,...,8$, or a two-component vector $\bm{M}_i$ for $i=8,...,12$. The polarization at order $\alpha^2$  
reads
\begin{eqnarray}
&& \Pi^{1}_i = -4g_i N_\psi^2 \alpha^2 \displaystyle\sum_{\eta,j} \displaystyle\int_{\bm{q},\omega} \text{Tr}[\bm{M}_i G_0'(\bm{q},\omega) T_j^{\bar{\eta}} G_0'(\bm{q}+\eta\bm{q}_j,\omega) \nonumber \\
&& \ \ \ \ \ \ \cdot \ \bm{M}_i G_0'(\bm{q}+\eta\bm{q}_j,\omega) T_j^{\eta} G_0'(\bm{q},\omega)].
\label{eqS:polari_2}
\end{eqnarray}
The pole of the integral in Eq.~\eqref{eqS:polari_1} per number of fermion flavors $n$ (equal to four in our case), is given by
 $\Pi_i^{0} = -4ng_i I_i/v\epsilon$ where
\begin{eqnarray}
&& I_i = \lim\limits_{\epsilon \to 0} \dfrac{v \epsilon N_\psi^2}{n} \displaystyle\int_{\bm{q},\omega}  \text{Tr}[\bm{M}_i G_0'(\bm{q},\omega) \cdot \bm{M}_i G_0'(\bm{q},\omega)] \nonumber\\
&& = \left\{
                \begin{array}{lcl}
\frac{1}{2\pi} & \text{if $\bm{M}_i$ has the sublattice structure} & \sigma_z, \\[0.2cm]
\frac{1}{4\pi} & \text{--} & \bm{\sigma}.
\end{array}
\right. \nonumber \\
\end{eqnarray}
In Eq.~\eqref{eqS:polari_2} we can use again the separation of energy scales : the theory is meaningful only at low energy, i.e. for $q, \omega \ll 1$, so that $G_0'(\bm{q}+\eta\bm{q}_j,\omega)$ can be replaced by $G_0'(\eta\bm{q}_j,0)$. This results in 
$\Pi_i^{1} = -3\alpha^2 \chi_i h_i(\beta) \Pi_i^{0}$ 
where $\chi_i$ equals either $+1$ for the interaction matrices $\sigma_z \tau_0$ and $\bm{\sigma}\tau_z$ or $-1$ for the interaction matrices $\sigma_z \tau_z$ and $\bm{\sigma} \tau_0$; and the corrugation-dependent function $h_i(\beta)$ equals either $1-\beta^2$ or $1$ if the interaction matrix matches $\sigma_0$ or $\sigma_z$ in the pseudospin sector, respectively. 
This fixes the renormalization constant to
\begin{equation}
\label{eqS:Z_phii}
Z_{\phi_i} = 1-\dfrac{4 n g_i I_i[1+ 3\alpha^2 \chi_i h_i(\beta)]}{v \epsilon}.
\end{equation}

\subsection{Vertices}

We denote the one-loop contribution to the three-point vertex of interaction $i$ renormalised by interaction $l$  by $V_{il}$. The one-loop vertices at zero external momentum $\bm{k}=\bm{0}$ and fixed frequency $\Omega$ are drawn in Fig.~\ref{figS:vertex_0} to \ref{figS:vertex_cro} and computed in Eq.~\eqref{eqS:vertex_0} to \eqref{eqS:vertex_4}. For the vertices correcting an interaction $i$ associated to a $2$d channel, we write the vertex for only one component of the matrix $\bm{M}_i$, simply denoted as $M_i$. The three-point vertex at order $\alpha^0$, given by diagram shown in Fig.~\ref{figS:vertex_0}, reads
\begin{widetext}
\begin{equation}
V^{0}_{il} = N_\psi^3 (2\sqrt{g_i})(4g_l) \displaystyle\int_{\bm{q},\omega} \bm{M_l} G_0'(\bm{q},\omega) M_i G_0'(\bm{q},\omega) \cdot \bm{M_l}.
\label{eqS:vertex_0}
\end{equation}
The three-point vertex at order  $\alpha^2$ (mixed diagram with two interlayer hopping) have either multiplicity one, or two. Those with multiplicity one nest either an internal hopping line, as in Fig.~\ref{figS:vertex_int},

\begin{equation}
V^{1,\text{int}}_{il} = N_\psi^3 (2\sqrt{g_i})(4g_l \alpha^2) \sum_{\eta,j} \int_{\bm{q},\omega} \bm{M_l} G_0'(\bm{q},\omega) T_j^{\bar{\eta}} G_0'(\bm{q}+\eta\bm{q}_j,\omega) M_i G_0'(\bm{q}+\eta\bm{q}_j,\omega) T_j^{\eta} G_0'(\bm{q},\omega) \cdot \bm{M_l},
\label{eqS:vertex_1}
\end{equation}
or an external hopping line, as in Fig.~\ref{figS:vertex_ext},
\begin{equation}
V^{1,\text{ext}}_{il} = N_\psi^3 (2\sqrt{g_i})(4g_l \alpha^2) \sum_{\eta,j} \int_{\bm{q},\omega} T_j^{\bar{\eta}} G_0'(\eta \bm{q}_j,\omega) \bm{M_l} G_0'(\bm{q},\omega) M_i G_0'(\bm{q},\omega) \cdot \bm{M_l} G_0'(\eta \bm{q}_j,\omega) T_j^{\eta}.
\label{eqS:vertex_2}
\end{equation}
The mixed diagrams with multiplicity two nest either an isolated hopping line, as in Fig.~\ref{figS:vertex_iso},
\begin{equation}
V^{1,\text{iso}}_{il}= 2N_\psi^3 (2\sqrt{g_i})(4g_l \alpha^2) \sum_{\eta,j} \displaystyle\int_{\bm{q},\omega} T_j^{\bar{\eta}} G_0'(\eta\bm{q}_j,\omega) \bm{M_l} G_0'(\bm{q}+\eta\bm{q}_j,\omega) T_j^{\eta} G_0'(\bm{q},\omega) M_i G_0'(\bm{q},\omega) \cdot \bm{M_l},
\label{eqS:vertex_3}
\end{equation}
or a hopping line that crosses the interaction line, shown in Fig.~\ref{figS:vertex_cro},
\begin{equation}
V^{1,\text{cro}}_{il} = 2N_\psi^3 (2\sqrt{g_i})(4g_l \alpha^2) \sum_{\eta,j} \int_{\bm{q},\omega} \bm{M_l} G_0'(\bar{\eta}\bm{q}_j, \omega) T_j^{\bar{\eta}} G_0'(\bm{q},\omega) M_i G_0'(\bm{q},\omega) \cdot \bm{M_l} G_0'(\eta\bm{q}_j, \omega) T_j^{\eta}.
\label{eqS:vertex_4}
\end{equation}
Similarly, we can define the pole of the integral appearing in Eq.~\eqref{eqS:vertex_0} as
\begin{equation}
J_{il} = \lim\limits_{\epsilon \to 0} \dfrac{v \epsilon N_\psi^2}{n} \int_{\bm{q},\omega} \text{Tr}[\bm{M_l} G_0'(\bm{q},\omega) M_i G_0'(\bm{q},\omega) \cdot \bm{M_l} M_i] = \left\{
                \begin{array}{lcl}
0 &  \text{if $(\bm{M}_i,\bm{M_l})$ match} & (\bm{\sigma},\bm{\sigma}), \\[0.2cm]
\frac{-1}{4\pi} &  \text{--} & (\bm{\sigma},\sigma_z), \\[0.2cm]
\frac{-1}{2\pi} & \text{--} & (\sigma_z,\bm{\sigma}), \\[0.2cm]
\frac{1}{2\pi} & \text{--} & (\sigma_z,\sigma_z).
\end{array}
\right. 
\end{equation}
such that $V^{1,\rm in}_{il} = V^{1,\rm ext}_{il} = 3\alpha^2 \chi_i
h_i(\beta) V^{0}_{il}$ with $V^{0}_{il} = N_\psi (2i\sqrt{g_i})(4g_l
J_{il}/v\epsilon)$. 
We also define the pole appearing in the sum of the diagrams with multiplicity two as

\begin{table*}[t!]
\begin{center}
\renewcommand{\arraystretch}{2}
\setlength\tabcolsep{0.55cm}
\begin{tabularx}{\textwidth-0\columnsep}{ccccc}
\hline
\hline
$f_{il}$ & $a_2^-~(\sigma_z\tau_0)$ & $a_1^-~(\sigma_z\tau_z)$ & $E_2^+~\left(\bm{\sigma} \tau_0/\sqrt{2}\right)$ & $E_4^- \left(\bm{\sigma} \tau_z/\sqrt{2}\right)$\\
\hline
~$a_2^-$ & $\frac{4}{\pi}\left[1 - 12\alpha^2(1-\beta^2)\right]$ & $-\frac{4 }{\pi} \left[1 - 6\alpha ^2(1-\beta^2)\right]$ & $\frac{4 }{\pi} \left[1-6 \alpha ^2(1+\beta ^2)\right]$ & $\frac{4 }{\pi} \left[1+6\alpha^2(3-\beta ^2)\right]$ \\
~$a_1^-$ & $-\frac{4}{\pi} \left[1 - 6\alpha ^2(1-\beta ^2)\right]$ & $\frac{4}{\pi}$ & $\frac{4}{\pi} \left[1-6\alpha ^2(1-\beta^2)\right]$ & $\frac{4}{\pi} \left[1-6\alpha^2(1-\beta ^2 )\right]$ \\
~$E_2^+$  & $\frac{2 }{\pi } \left[1-6 \alpha ^2(1+\beta^2)\right]$ & $\frac{2}{\pi} \left[1-6 \alpha ^2(1-\beta^2) \right]$ & $\frac{4}{\pi} \left[1-3 \alpha^2(1-\beta^2)\right]$ & $-\frac{12}{\pi} \alpha^2 \beta^2$ \\
~$E_4^-$ & $\frac{2 }{\pi } \left[1+6\alpha ^2(3-\beta^2)\right]$ & $\frac{2 }{\pi } \left[1-6\alpha^2(1-\beta^2)\right]$ & $-\frac{12}{\pi} \alpha^2 \beta^2$ & $\frac{4}{\pi} \left[1 + 3 \alpha ^2(1+\beta^2)\right]$ \\
\hline
\hline
\end{tabularx}
\label{tab:characters_D3C}
\end{center}
\caption{List of the functions $f_{il}(\alpha,\beta)$ appearing in the RG flows of the four non-trivial channels $a_2^-$, $a_1^-$ $E_2^+$, and $E_4^-$. The interaction matrices associated to each of these channels are indicated in the first line of the table.}
\end{table*}

\begin{multline}
K_{il}(\beta) = \\
\lim\limits_{\epsilon \to 0} \dfrac{v \epsilon N_\psi^4}{n} \sum_{\eta,j} \displaystyle\int_{\bm{q},\omega} \text{Tr}[\bm{M_l} G_0'(\eta\bm{q}_j,\omega) T_j^{\eta} G_0'(\bm{q},\omega) M_i G_0'(\bm{q},\omega) \cdot \bm{M_l} (M_i T_j^{\bar{\eta}} G_0'(\eta\bm{q}_j,\omega) + G_0'(\bar{\eta}\bm{q}_j, \omega) T_j^{\bar{\eta}} M_i)],
\end{multline}
\end{widetext}
such that $V^{1,\rm iso}_{il} + V^{1,\rm cro}_{il} = N_\psi (2i\sqrt{g_i})(8\alpha^2 g_l K_{il}(\beta)/v\epsilon)$. 
The integrals $J_{il}$ are numerical constants, dependent of neither the number of fermion flavors $n$ nor the corrugation parameter $\beta$, while $K_{il}(\beta)$ depends on the corrugation parameter. Using commutation relations between interaction and hopping matrices, we can express all vertices \eqref{eqS:vertex_0} -- \eqref{eqS:vertex_4} in terms of $J_{il}$ and $K_{il}$ only. We then find the vertex renormalization constant to be
\begin{equation}
\label{eqS:Z_gi}
Z_{g_i} = 1-\dfrac{4}{v \epsilon} \sum_l  g_l [(1+ 6\alpha^2  h_i(\beta) \chi_i) J_{il} + 2 \alpha^2 K_{il}(\beta)].
\end{equation}

\subsection{RG  flow equations}
We express   $Z_i = Z_{g_i}^2 Z_{\phi_i}^{-1}$ to first order in the coupling constants as
\begin{equation}
Z_i =  1 +  \sum_{l=1}^{12} \dfrac{f_{il}(\alpha,\beta) g_l}{v \epsilon},
\end{equation}
where 
\begin{eqnarray}
f_{il}(\alpha,\beta) &=& 4\left[ (nI_i\delta_{il} - 2 J_{il}) \right.  + 3\alpha^2 h_i(\beta) \chi_i(nI_i\delta_{il} \nonumber\\
&& \left. - 4 J_{il}) - 4 \alpha^2 K_{il}(\beta)) \right]. \label{eqS:fij_1D}
\end{eqnarray}
and $v$ is the Fermi velocity~\eqref{eqS:fermi_velocity}.  
We compute the RG flow equations by deriving Eq.~\eqref{eqS:Z_couplings} with respect to $\mu$ at constant bare couplings. This yields 
\begin{equation}\label{eqS:beta_func}
- \dfrac{\partial \log g_i}{\partial \log \mu} = -\epsilon  + v^{-1}\sum_{l=1}^{12} f_{il}(\alpha,\beta) g_l.
\end{equation}


\newpage

\bibliographystyle{apsrev4-1}
\bibliography{biblioTBG}

\end{document}